\documentclass[11pt]{article}%

\usepackage{amsmath,amssymb,amsfonts}
\usepackage{calc}
\usepackage{dsfont}
\usepackage{slashed}
\usepackage{cite}
\usepackage{epsfig}
\usepackage{slashed}
\usepackage{multirow}
\usepackage[bottom]{footmisc}

\renewcommand{\arraystretch}{1.1}
\newcommand{\beq}{\begin{eqnarray}}
\newcommand{\eeq}{\end{eqnarray}}
\newcommand{\gsim}{\raisebox{-0.13cm}{~\shortstack{$>$ \\[-0.07cm]
      $\sim$}}~}
\newcommand{\lsim}{\raisebox{-0.13cm}{~\shortstack{$<$ \\[-0.07cm]
      $\sim$}}~}
\newcommand{\s}{\newline \vspace*{-3.5mm}}

\makeatletter
\def\section{\@startsection {section}{1}{\z@}{-3.5ex plus -1ex minus -.2ex}{2.3ex plus .2ex}{\large\bf}}
\def\subsection{\@startsection{subsection}{2}{\z@}{-3.25ex plus -1ex
minus -.2ex}{1.5ex plus .2ex}{\normalsize\bf}}
\makeatother
\newcommand{\captionfonts}{\small}
\makeatletter  
\long\def\@makecaption#1#2{%
  \vskip\abovecaptionskip
  \sbox\@tempboxa{{\captionfonts #1: #2}}%
  \ifdim \wd\@tempboxa >\hsize
    {\captionfonts #1: #2\par}
  \else
    \hbox to\hsize{\hfil\box\@tempboxa\hfil}%
  \fi
  \vskip\belowcaptionskip}
\makeatother   

\graphicspath{{fig/}}

\begin{document}

\thispagestyle{empty}

\begin{center}

\hfill KA-TP-38-2013 \\
\hfill SFB/CPP-13-97 \\
\hfill CP3-Origins-2013-041 \\
\hfill DIAS-2013-41 \\

\begin{center}

\vspace{1.7cm}

{\LARGE\bf Vector-like Bottom Quarks in Composite Higgs Models} 
\end{center}

\vspace{1.4cm}

{\bf M.~Gillioz$^{\,a}$}, {\bf R.~Gr\"ober$^{\,b}$}, 
{\bf A.~Kapuvari$^{\,b}$} and {\bf M.~M\"uhlleitner$^{\,b}$} \\

\vspace{1.2cm}

${}^a\!\!$
{\em {CP$^3$-Origins and Danish Institute for Advanced Study,
University of Southern Denmark, Campusvej 55, DK-5230, Odense M, Denmark and}
}\\
${}^b\!\!$
{\em {Institut f\"ur Theoretische Physik, Karlsruhe Institute of Technology, D-76128 Karlsruhe, Germany}
}

\end{center}

\vspace{0.8cm}

\centerline{\bf Abstract}
\vspace{2 mm}
\begin{quote}
\small
Like many other models, Composite Higgs Models feature the existence
of heavy vector-like quarks. Mixing effects between the Standard Model
fields and the heavy states, which can be quite large in case of the
top quark, imply deviations from the SM. In this work we investigate
the possibility of heavy bottom partners. We show that 
they can have a significant impact on electroweak precision
observables and the current Higgs results if there is a sizeable
mixing with the bottom quark. We explicitly check that the
constraints from the measurement of the CKM matrix element $V_{tb}$
are fulfilled, and we test the compatibility with the electroweak precision
observables. In particular we evaluate the constraint
from the $Z$ coupling to left-handed bottom quarks. General
formulae have been derived which include the effects of new bottom
partners in the loop corrections to this coupling and which can be
applied to other models with similar particle content. Furthermore,
the constraints from direct searches for heavy states at the LHC and
from the Higgs search results have been included in our analysis. The
best agreement with all the considered constraints is achieved for medium to
large
compositeness of the left-handed top and bottom quarks.
\end{quote}

\newpage
\renewcommand{\thefootnote}{\arabic{footnote}}
\setcounter{footnote}{0}

\section{Introduction}
The announcement of the discovery of a new scalar particle by
the LHC experiments ATLAS \cite{:2012gk} and CMS \cite{{:2012gu}} has
marked a milestone in elementary particle physics. Since then, the
properties of the particle have been investigated and strongly suggest
it to be the Higgs boson, {\it i.e.}~the particle related to the Higgs
mechanism. So far no new additional particles have been discovered
which could help to clarify the question which is the dynamics behind
electroweak symmetry breaking (EWSB). It could be weakly interacting
like in the Standard Model (SM) or in its supersymmetric
extensions. The Higgs particle could also arise as pseudo
Nambu-Goldstone boson (pNGB) from 
a strongly-coupled sector \cite{strong1,strong2}, as is the case in
Composite Higgs Models. In the Strongly-Interacting
Light Higgs (SILH) \cite{Giudice:2007fh} scenario there exists a light, narrow
Higgs-like scalar, which is a bound state from some strong dynamics. 
Due to its Goldstone nature, the Higgs boson is separated from the other usual
resonances of the strong sector by a mass gap. 
The low-energy particle content is the same as in the
SM. In Composite Higgs Models the problem of the fermion mass
generation is solved by the idea of partial compositeness \cite{partial}. The SM
fermions are elementary particles which couple linearly to the heavy
states of the strong sector that carry equal quantum numbers. 
In particular the top quark can be largely composite. The linear
couplings of the SM particles to the strong sector explicitly break the
global symmetry of the latter
and the Higgs potential arises from loops of SM particles, with
the top quark giving the main contribution.
In order to naturally accommodate a
low-mass Higgs boson of $\sim 125$~GeV the top partners should be rather
light, with masses~$\lsim 1$~TeV
\cite{Contino:2006qr,lighttops,MarzoccaSerone,Pomarol:2012qf},
depending on the model and the scale of compositeness. This bound can
be relaxed somewhat by 
contributions from new heavy gluons \cite{Barnard:2013hka}. 
Heavy vector-like resonances in this mass range can be produced and searched
for at the LHC \cite{vlsearch,Bini:2011zb,Andeen}. \s

The SILH
\cite{Giudice:2007fh} Lagrangian arises as first
term of an expansion in $\xi = v^2/f^2 \ll 1$,
where $v$ is the scale of EWSB and $f$ is the typical scale of the
strong sector. It can be used in the vicinity of
the SM limit given by $\xi \to 0$. For larger values of $\xi$ a
resummation of the series in $\xi$ is required. Explicit models built
in five-dimensional warped space can provide such a resummation. In the
Minimal Composite Higgs Model (MCHM) of Ref.~\cite{Agashe:2004rs},
which is based on a 5-dimensional theory in Anti-de-Sitter space-time,
the bulk symmetry $SO(5)\times U(1)_X \times SU(3)$ is broken down to
the SM gauge group on the ultraviolet (UV) boundary and to $SO(4)
\times U(1)_X \times SU(3)$ on the infrared (IR). The mixing effects
between the SM fields and the heavy states of the new 
sector, which arise at tree-level, lead to sizable deviations from the
SM predictions. Composite Higgs Models are therefore mainly challenged
by the electroweak precision tests (EWPT)
\cite{Agashe:2005dk,Barbieri:2007bh, Pomarol:2008bh}. Particularly strong
constraints 
can arise from the $Zb_L\bar{b}_L$ coupling, which has been measured 
very precisely and agrees with the SM prediction at the sub-percent
level. With the top quark mixing strongly with the new sector, the
left-handed bottom quark $b_L$ which is in the same weak doublet as
$t_L$ receives large modifications of its couplings. The $Zb_L
\bar{b}_L$ coupling is safe from large corrections if the fermions are
embedded in fundamental {\bf 5} or {\bf 10} representations of
$SO(5)$, where $b_L$ belongs to a bi-doublet ({\bf
2},{\bf 2}) of $SU(2)_L \times SU(2)_R$, and the $SO(4)$ symmetry is
enlarged to $O(4)$ \cite{Agashe:2006at}. Subsequent investigations
including the fermion composites in full representations of the
$SO(5)$ \cite{Lodone:2008yy,Gillioz:2008hs} and extended to models
with multiple sets of fermionic composites \cite{Anastasiou:2009rv}
showed that Composite Higgs Models can fulfill the constraints of EWSB. 
Further constraints on these models come from flavour
physics. Four-fermion operators that arise in Composite Higgs Models
contribute to flavor-changing processes and electric dipole
moments. The flavour structure of the strong sector cannot be
predicted through naturalness considerations, and a variety of flavour
implementations can be realized
\cite{anarchic,Csaki:2008zd,u1h3,u2approx,Redi:2011zi,Redi:2012uj,flavourconstraints,DaRold:2012sz,Barbieri:2012tu}. \s

The Composite Higgs couplings to the SM particles are changed with respect to the ones of the SM Higgs boson. In the MCHMs of
Refs.~\cite{Contino:2006qr,Agashe:2004rs,Contino:2003ve} they can be 
parametrized in terms of a single parameter $\xi$. These coupling
modifications change the Higgs boson phenomenology
\cite{Low:2009di,Contino:2010mh,Espinosa:2010vn,Low:2010mr,Grober:2010yv,Espinosa:2012qj,Azatov:2012qz,Montull:2013mla,Azatov:2013ura,Contino:2013gna}. 
With the top quark being a composite particle, mixing effects with the
heavy top partners induce further changes in the top-Higgs Yukawa
coupling. In  addition top partners running in the loops of the loop-induced 
Higgs couplings to gluons and photons could lead to sizeable
corrections of these vertices. It has been shown
\cite{Falkowski:2007hz,Low:2010mr,Montull:2013mla,Azatov:2011qy,Gillioz:2012se,Delaunay:2013iia},
however, that these vertices depend on the pure non-linearities of the
Higgs boson and are not sensitive to the details of the resonance
spectrum. By applying the low-energy theorem
(LET)\cite{Kniehl:1995tn}, it can be shown that the corrections to the
Yukawa coupling and the contributions from the extra fermion loops
cancel each other, so that the loop-induced couplings only depend on
$v/f$. The bottom quark, being the next-heaviest quark, 
implies a sizable mixing with the strong sector also for the bottom.
In this case, due to the small bottom
mass, the LET cannot be applied any more and the Higgs 
loop-couplings to gluons and photons will depend on the resonance
structure of the strong sector, with significant implications for the Higgs
phenomenology \cite{Azatov:2011qy, Delaunay:2013iia, Barducci:2013wjc}. \s

The aim of this paper is to study the implications of composite bottom
quarks on the viability of Composite Higgs Models and on the LHC Higgs
phenomenology by introducing a minimum amount of new parameters. 
For this purpose the fermions are
embedded in the ${\bf 10}$, which is the smallest possible
representation of $SO(5)$ that allows to include partial compositeness
for the bottom quarks, while being compatible with the EWPTs by
implementing custodial symmetry. 
The outline of the paper is as follows. In section
\ref{sec:model} we present the 
model. In section \ref{sec:ewpt} the new contributions to the
electroweak precision observables due to the composite nature of the
$b$-quark and to the additional heavy resonances are computed, in
particular the new contributions to the loop corrections of the $Zb_L
\bar{b}_L$ coupling. We then perform a $\chi^2$ test to investigate the
compatibility of the model with the constraints that arise from
electroweak precision measurements. Section \ref{sec:higgsres} is
devoted to the constraints from the LHC Higgs results and the
direct searches for heavy fermions. In order to compare with the
experimental best fit values to the Higgs rates, the Higgs production
and decay processes are calculated for the model. Likewise the
mass spectrum of the heavy fermion sector and the decay widths of the
new resonances are computed and confronted to the LHC
searches for heavy fermions. A brief discussion on implications from
flavour physics is 
included. In section \ref{sec:numerical} we present our numerical
results. The $\chi^2$ test taking into account the EWPTs and the newest experimental measurement of the CKM
matrix element $V_{tb}$ is extended to include the latest
Higgs rates reported by the experiments. Our results are summarized in
the conclusions, section \ref{sec:conclusion}.


\section{The Model \label{sec:model}}
The models given in Refs.~\cite{Agashe:2004rs,Contino:2006qr} have
been constructed in terms of five-dimensional theories on Anti-de-Sitter
space-time and provide a resummation for large values of $\xi$.  In
the following we will work in the simplest model including
custodial symmetry and allowing for the inclusion of bottom quarks as
composite objects. We will show the 
effects of heavy bottom partners for a minimal $SO(5)\times
U(1)_X/SO(4)\times U(1)_X$ symmetry breaking pattern, where the additional
$U(1)_X$ is introduced to guarantee the correct fermion charges. The
electroweak group $SU(2)_L \times U(1)_Y$ of the SM is embedded into
$SO(4) \times U(1)_X$ with the hypercharge $Y$ given by $Y=T^3_R
+X$. The coset $SO(5)/SO(4)$ provides four Goldstone bosons, three of
them are the longitudinal modes of the vector bosons and one is the
Higgs boson. The four Goldstone bosons can be parameterized in terms
of the field 
\begin{equation}
 \Sigma=\Sigma_0 \exp(\Pi(x)/f), \hspace*{1cm}
\Sigma_0=(0,0,0,0,1)\;,\hspace*{1cm}\Pi(x)=-i\sqrt{2}
T^{\hat{a}}h^{\hat{a}}(x)\;,
\end{equation}
with $T^{\hat{a}}$ ($\hat{a}=1..4$) denoting the generators of the coset
$SO(5)/SO(4)$. 
They are given by
\beq
(T^{\hat{a}})_{ij}&=-\frac{i}{\sqrt{2}}\left(\delta^{\hat{a}}
_i\delta^5_j-\delta^{\hat{a}}_j\delta^5_i\right)\;.
\eeq
Together with the generators of the $SU(2)_{L,R}$ 
$(a,b,c=1,2,3, \, i,j=1,...,5)$,
\begin{align}
 (T^{a}_{L})_{ij}&=-\frac{i}{2}\left[\frac{1}{2}\epsilon^{abc}
(\delta^b_i\delta^c_j-\delta^b_j\delta^c_i)+\delta^a_i\delta^4_j-\delta^4_i
\delta ^a_j\right]\;,\label{mod01}
\\
(T^{a}_{R})_{ij}&=-\frac{i}{2}\left[\frac{1}{2}\epsilon^{abc}
(\delta^b_i\delta^c_j-\delta^b_j\delta^c_i)-\delta^a_i\delta^4_j+\delta^4_i
\delta ^a_j\right]\;,
\end{align}
they form the generators for the fundamental representation of $SO(5)$.
%
This leads to the explicit expression for the Goldstone field
$\Sigma$,
\begin{equation}
 \Sigma=\frac{\sin h/f}{h}(h_1, h_2, h_3, h_4, h\cot(h/f))\; ,\hspace*{1cm}
h=\sqrt{\sum_{\hat{a}=1}^4 h_{\hat{a}}^2}\;.\label{mod10}
\end{equation}
The low-energy physics of the strong sector can be described by a non-linear
$\sigma$-model. The kinetic term of the Goldstone field can then be written as
\begin{equation}
 \mathcal{L}_{kin}=\frac{f^2}{2}\left(D_{\mu}\Sigma\right)\left(D^{\mu}
\Sigma\right)^T\;, \quad \mbox{with} \quad
D_{\mu}\Sigma=\partial_{\mu}\Sigma -i g' B_{\mu} \Sigma(T^3_R+X)-i g W_{\mu}^a
\Sigma T^a_L\;\label{mod06},
\end{equation}
where $W_\mu^a$ and $B_\mu$ are the electroweak $SU(2)$ and $U(1)$
fields, respectively, with the corresponding couplings $g$ and $g'$.
In the unitary gauge the vacuum expectation value (VEV) can be aligned
with the direction of $h_4$ which is identified with $H \equiv h_4$,
so that
\begin{equation}
 \Sigma=(0,0,0,\sin(H/f),\cos(H/f))\;,
\end{equation}
and we get for the kinetic term 
\begin{equation}
\mathcal{L}_{kin}=\frac{1}{2}\partial_{\mu}H\partial^{\mu}H+\frac{f^2}{4}
\sin^2\left(\frac{H}{f}\right)\left[g^2
W_{\mu}^+W^{\mu-}+\frac{g^2}{2\cos^2\theta_W}Z^{\mu}Z_{\mu}\right] \;.
\label{mod07} 
 \end{equation}
Expanding Eq.~(\ref{mod07}) in powers of the Higgs field $H=\langle H
\rangle +h$, and identifying 
\begin{equation}
\xi = \left( \frac{v}{f} \right)^2 =\sin^2 \frac{\langle H\rangle}{f} \;,
\label{eq:xidef}
\end{equation}
one obtains
the couplings to the gauge fields in terms of the corresponding SM
couplings ($V=W,Z$)
\beq
g_{hVV} = g_{hVV}^{\text{SM}} \, \sqrt{1-\xi} \;, \qquad g_{hhVV} =
g_{hhVV}^{\text{SM}} \, (1-2\xi)\;, \label{eq:higgsv}
\eeq
and the usual mass relation $m_W^2=g^2v^2/4$, with
$v=1/\sqrt{G_F\sqrt{2}} \approx 246$~GeV. \s

New fermionic resonances in Composite Higgs Models are expected to be well
below the cut-off of the effective theory in order to accommodate a
Higgs boson with mass $m_h\approx 125$ GeV
\cite{lighttops,MarzoccaSerone,Pomarol:2012qf}. 
Fermion mass generation is then achieved by the principle of partial
compositeness, in which an elementary fermion acquires its mass through
the mixing with new vector-like fermions of the strong sector. 
This can be implemented in the Lagrangian through linear couplings of the
elementary sector with the strong sector. The quantum numbers of the
new fermion must be such that the Lagrangian is invariant under the SM
gauge group. A large, phenomenologically interesting mixing occurs if
the SM fermion is heavy, which makes the discussion of the
third generation quarks the most interesting.\footnote{Partial
  compositeness of the light quarks has been discussed
  in~\cite{Delaunay:2013iia, lightflavour2} and of the leptons
  in~\cite{Carmona:2013cq}.} Previous works, as {\it e.g.}
Refs.~\cite{Anastasiou:2009rv,Gillioz:2008hs,Gillioz:2012se}, have
studied the mass generation of the top quark through mixing, while the
bottom quark was taken massless or introduced {\it
  ad hoc}. 
The purpose of this work, however, is to study the effect 
of bottom partners that arise when the bottom quark mass is generated
by mixing with the strong sector. This cannot be achieved by introducing
only a single fermion multiplet in the fundamental or spinorial
representation of $SO(5)$.  In the following we will therefore
consider a {\bf $\text{\bf 10}_{2/3}$}, which is the smallest
representation of $SO(5)$ having the desired features. Note that since
there is only one multiplet giving a mass both to the top and bottom
quark, no new parameters need to be introduced compared to the
previous models where a {\bf $\text{\bf 5}_{2/3}$} is used to generate
a mass for the top quark. If there are no new resonances of the strong
sector below the cut-off, apart from the Higgs boson, the model
displays the same phenomenology as the one with fermions embedded in
the fundamental representation, {\it
  cf.}~Ref.~\cite{Contino:2006qr}. The {\bf 10} of $SO(5)$ is a
two-index  antisymmetric representation, which can be written as follows 
\begin{equation}
\begin{split}
&\mathcal{Q}=\frac{1}{2}\\
&\begin{pmatrix}0&-(u+u_1)&\frac{i(d-\chi)}{\sqrt{2}}+
\frac{i(d_1-\chi_1)}{\sqrt{2}}&\frac{d+\chi}{\sqrt{2}}-
\frac{d_1+\chi_1}{\sqrt{2}}&d_4+\chi_4\\
u_1+u&0&\frac{d_1+\chi_1}{\sqrt{2}}+\frac{d+\chi}{\sqrt{2}}&
\frac{i(d_1-\chi_1)}{\sqrt{2}}-\frac{i(d-\chi)}{\sqrt{2}}&-i(d_4-\chi_4)\\
 -\frac{i(d_1-\chi_1)}{\sqrt{2}}-\frac{i(d-\chi)}{\sqrt{2}}&
-\frac{d_1+\chi_1}{\sqrt{2}}-\frac{d+\chi}{\sqrt{2}}
&0&u_1-u&t_4+T_4\\
 \frac{d_1+\chi_1}{\sqrt{2}}-\frac{d+\chi}{\sqrt{2}}&\frac{i(\chi_1-d_1)
} { \sqrt {2}}+\frac{i(d-\chi)}{\sqrt{2}}
&u-u_1&0&-i(t_4-T_4)\\
 -d_4-\chi_4&i(d_4-\chi_4)&-t_4-T_4&i(t_4-T_4)&0\end{pmatrix}\end{split}
\end{equation}
where the fermions $u, u_1, t_4$ and $T_4$ have electric charge 2/3,
$d, d_1$ and $d_4$ have charge -1/3, and  the charge of 
$\chi, \chi_1$ and $\chi_4$ is 5/3.
\begin{table}\centering
\renewcommand*{\arraystretch}{1.2}
\begin{tabular}{|c||c|c|c|c|c|c|c|c|c|c|c|}
\hline &$u$&$u_1$&$t_4$&$T_4$&$d$&$d_1$&$d_4$&$\chi$&$\chi_1$&$\chi_4$\\ \hline\hline
$T_{3,L}$&0&0&-1/2&1/2&-1&0&-1/2&1&0&1/2\\\hline
$T^2_L$&1&0&1/2&1/2&1&0&1/2&1&0&1/2\\\hline
$T_{3,R}$&0&0&1/2&-1/2&0&-1&-1/2&0&1&1/2\\\hline
$T^2_R$&0&1&1/2&1/2&0&1&1/2&0&1&1/2\\\hline
$Y$&2/3&2/3&7/6&1/6&2/3&-1/3&1/6&2/3&5/3&7/6\\\hline
\end{tabular}
\caption{Quantum numbers of the new vector-like fermions under
  $SU(2)_L \times SU(2)_R$. The last line is the hypercharge.}\label{tab1}
\end{table}
The decomposition of the {\bf 10} under $SU(2)_L \times SU(2)_R$ is
\begin{equation}
 \mathbf{10}=(\mathbf{2,2}) \oplus (\mathbf{3,1}) \oplus (\mathbf{1,3})\;.
\end{equation}
The exact quantum numbers of each of the new fermions can be read off
Table~\ref{tab1}. The Lagrangian including the new fermion multiplet
$\mathcal{Q}$ then reads
\begin{equation}
\begin{split}
 \mathcal{L}=& \, i\,\text{Tr}(\bar{\mathcal{Q}}_R\slashed{D}\mathcal{Q}_R)+
i\,\text{Tr}(\bar{\mathcal{Q}}_L\slashed{D}\mathcal{Q}_L)+i\bar{q}_{L}
\slashed{D}q_{L}+i\bar{b}_{R}\slashed{D}b_{R}+i\bar{t}_{R}
\slashed{D}t_{R}\\&-M_{10} \text{Tr}(\bar{\mathcal{Q}}_R\mathcal{Q}_L)-y
 f\left(\Sigma^\dagger \bar{\mathcal{Q}}_R\mathcal{Q}_L \Sigma\right)+h.c.
\\&-\lambda_t\bar{t}_R u_{1L}-\lambda_b \bar{b}_R d_{1L}
-\lambda_q (\bar{T}_{4R}, \bar{d}_{4R})q_L+h.c.\label{lagragian}\;,
\end{split}
\end{equation}
where the SM doublet of the left-handed top and bottom quark is
denoted by $q_L$ and the right-handed top (bottom) quark by $t_R$
($b_R$). The covariant derivative acts on $\mathcal{Q}$ as
 \begin{equation}
D_{\mu}\mathcal{Q}=\partial_{\mu}\mathcal{Q}-i g W^{a}[T^a_L,\mathcal{Q}]-ig' B
\big( [T^3_R, \mathcal{Q}]+X\mathcal{Q} \big)\;,\hspace*{1cm} X=(2/3)
\mathds{1}\;,\label{mod08}
\end{equation}
with the generators $T^a_L$ defined as in Eq.~\eqref{mod01}. Note that the
mixing terms with the coupling constants $\lambda_q$,$\lambda_t$ and
$\lambda_b$ explicitly break $SO(5)$. 
Using the abbreviations 
\beq 
s_H\equiv \sin(H/f) \;, \qquad c_H \equiv \cos(H/f)
\eeq
and
\begin{equation}
\tilde{m}_a \equiv\frac{1}{4} f y s_H^2+M_{10}\;,\hspace*{1cm}
\tilde{m}_b \equiv \frac{1}{2} f y(1-\frac{1}{2}s_H^2)+M_{10} \;,\hspace*{1cm}
\tilde{m}_c \equiv \frac{1}{2} f yc_H^2+M_{10}\;,
\end{equation}
the terms of the Lagrangian in Eq.~\eqref{lagragian}, which are
bilinear in the quark fields, can be written as
\small\begin{equation}
-\mathcal{L}_{m_{t}}=
\overline{\left(\begin{array}{c}
t_L\\ u_L\\u_{1L}\\t_{4L}\\T_{4L}
\end{array}\right)}
\left(
\begin{array}{ccccc}
 0 & 0 & 0 & 0 & \lambda_q \\
 0 & \tilde{m}_a& -\frac{1}{4} f
   y s_H^2 & -\frac{1}{4} f y
  c_H s_H & -\frac{1}{4} f y c_Hs_H
   \\
 \lambda_t & -\frac{1}{4} f y s_H^2 & \tilde{m}_a& \frac{1}{4}
f
   yc_H  s_H & \frac{1}{4} f y c_Hs_H
   \\
 0 & -\frac{1}{4} f yc_H s_H & \frac{1}{4} f y c_Hs_H
   & \tilde{m}_b& -\frac{1}{4} f
   y s_H^2 \\
 0 & -\frac{1}{4} f yc_H s_H & \frac{1}{4} f y c_Hs_H
   & -\frac{1}{4} f ys_H^2 &
   \tilde{m}_b
\end{array}
\right) 
\left(\begin{array}{c}
t_R\\ u_R\\u_{1R}\\t_{4R}\\T_{4R}
\end{array} \right)+h.c.\;,\label{app101}
\end{equation}

\vspace*{0.1cm}
\begin{equation}
-\mathcal{L}_{m_{b}}=\overline{ \left(\begin{array}{c}
                      b_L \\d_L \\ d_{1L}\\ d_{4L} 
                      \end{array}\right)}
\left(
\begin{array}{cccc}
 0 & 0 & 0 & \lambda_q \\
 0 &\tilde{m}_a& -\frac{1}{4} f
   ys_H^2 & f y \frac{c_H s_H}{2 \sqrt{2}} \\
 \lambda_b& -\frac{1}{4} f
   ys_H^2  &\tilde{m}_a& -f y
  \frac{c_H s_H}{2 \sqrt{2}} \\
 0 & f y\frac{c_H s_H}{2 \sqrt{2}} & -f y\frac{
  c_H s_H}{2 \sqrt{2}} & \tilde{m}_c
\end{array}
\right)
\left(\begin{array}{c}
                      b_R \\d_R \\ d_{1R}\\ d_{4R} 
                      \end{array}\right)+h.c.\;,\label{app102}
\end{equation}\normalsize

\vspace*{0.1cm}
\noindent
and
\small\begin{equation}
-\mathcal{L}_{m_{\chi}}= \overline{\left(\begin{array}{c}
                          \chi_L\\\chi_{1L}\\\chi_{4L}
                         \end{array}\right)}
\left(
\begin{array}{ccc}
\tilde{m}_a& -\frac{1}{4} f
   ys_H^2 & f y\frac{ c_H s_H}{2 \sqrt{2}} \\
 -\frac{1}{4} f ys_H^2 &
  \tilde{m}_a & -f y\frac{
  c_H s_H}{2 \sqrt{2}} \\
 f y\frac{c_H s_H}{2 \sqrt{2}} & -f y\frac{
  c_H s_H}{2 \sqrt{2}} & \tilde{m}_c
\end{array}
\right)
\left(\begin{array}{c}
                          \chi_R\\\chi_{1R}\\\chi_{4R}
                         \end{array}\right)+h.c.\;.\label{app103}
\end{equation}\normalsize

\vspace*{0.1cm}
\noindent
The mass matrices $M_{t}$, $M_{b}$ and $M_{\chi}$ can be obtained by
replacing the Higgs field in Eqs.~\eqref{app101}-\eqref{app103},
encoded in $s_H$ and $c_H$, respectively, with its VEV, 
{\it i.e.}~$H\to \langle H\rangle$. They are diagonalized by a bi-unitary
transformation
\begin{equation}
 \left(U_L^{(t/b/\chi)}\right)^{\dagger}M_{(t/b/\chi)}U_R^{
(t/b/\chi) } =M_ {(t/b/\chi)}^{diag}\;,\label{mod09}
\end{equation}
where $U_{L,R}^{(t/b/\chi)}$ denote the transformations that
diagonalize the mass matrix in the top, bottom and charge-5/3 ($\chi$)
sector, respectively.
In our analysis we diagonalize them numerically, setting the values of
$\lambda_t$ and
$\lambda_b$ such that the physics values of the top and bottom quark
masses are recovered. An analytic understanding of the size of the
masses can be obtained before electroweak symmetry breaking, {\it
  i.e.} for $v=0$. The following rotations diagonalize the mass
matrices
\beq 
\begin{array}{lllll}
\begin{pmatrix} q_{L} \\ Q_{L} \end{pmatrix} &\to& 
\begin{pmatrix} \cos\phi_{L} & \sin\phi_{L} \\
-\sin\phi_{L} & \cos\phi_{L} \end{pmatrix} & \begin{pmatrix}q_{L} \\ Q_{L}
\end{pmatrix} \, , & \hspace*{1cm}
\tan\phi_L=\lambda_q/(M_{10}+f y/2)\;, \\[0.4cm]
\begin{pmatrix} t_{R} \\ u_{1R} \end{pmatrix} &\to& \begin{pmatrix}
\cos\phi_{Rt} & \sin\phi_{Rt} \\
-\sin\phi_{Rt} & \cos\phi_{Rt} \end{pmatrix} & \begin{pmatrix}t_{R} \\
u_{1R} \end{pmatrix} \, , & \hspace*{1cm}
\tan\phi_{Rt}=\lambda_{t}/M_{10}\;,\\[0.4cm]
\begin{pmatrix} b_{R} \\ d_{1R} \end{pmatrix} &\to& \begin{pmatrix}
\cos\phi_{Rb} & \sin\phi_{Rb} \\
-\sin\phi_{Rb} & \cos\phi_{Rb} \end{pmatrix} & \begin{pmatrix}b_{R}
\\ d_{1R} \end{pmatrix} \, , & \hspace*{1cm}
\tan\phi_{Rb}=\lambda_{b}/M_{10}\;, 
\end{array}
\label{PCrotations}
\eeq
with $Q_{L}=(T_{4L}, d_{4L})$. The masses of the top partners are then
found to be 
\begin{equation} 
M_{10} \, ,\quad \frac{M_{10}}{\cos\phi_{Rt}} \, ,\quad M_{10}+\frac{f
  y}{2} \, ,\quad \frac{M_{10}+\frac{f y}{2}}{\cos\phi_L} \; , 
\label{eq:v0topmasses}
\end{equation}
and the masses of the bottom partners
\begin{equation}
 M_{10} \, , \quad \frac{M_{10}}{\cos\phi_{Rb}} \, , \quad
 \frac{M_{10}+\frac{f y}{2}}{\cos\phi_{L}}\;.
\label{eq:v0bottommasses}
\end{equation}
If the new scale $f$ is much larger than $v$ an expansion in $v/f$ of the mass
matrices can be performed. At leading order in $v/f$, this yields for
the top and bottom quark mass
\begin{equation}
  m_{top}=\frac{y\;v}{4}\sin\phi_L\sin\phi_{Rt}\;,
\hspace{1cm}m_{bot}=\frac{y\;v}{2\sqrt{2}}\sin\phi_L\sin\phi_{Rb}\;.
\label{mod13}
\end{equation}
We see, that in order to achieve the experimentally measured value of 
the top quark, $t_L$ and $t_R$ cannot be too elementary at the same 
time. Furthermore, as the top and bottom quark are in the same
doublet, the compositeness of the left-handed bottom is directly
connected to the compositeness of the left-handed top. As $\sin\phi_L$
cannot be too small in order to reproduce the top quark mass, this
implies that the right-handed component of the bottom quark is mostly
elementary, so that a small enough bottom mass can be achieved. 
The first correction term to the top and bottom partner masses is of
$\mathcal{O}(v^2/f^2) $.
For the charge-$5/3$ fermions the masses can be computed analytically
even for non-vanishing $v$. They are given by
\begin{equation}
 M_{10} \,,\quad M_{10}\, ,\quad M_{10}+\frac{f y}{2}\;.
\end{equation}
The Higgs coupling matrices can be obtained from
Eqs.~\eqref{app101}-\eqref{app103} by expanding the mass matrices in
the interaction eigenstates up to first order in the Higgs field
$H$. They read
\small\begin{equation} 
-\mathcal{L}_{h t\bar{t}}=
y\;h\overline{\left(\begin{array}{c}
t_L\\ u_L\\u_{1L}\\t_{4L}\\T_{4L}
\end{array}\right)}
\underbrace{\left(\begin{array}{ccccc}
 0 & 0 & 0 & 0 & 0 \\
 0 & \frac{1}{2}s_H c_H& - \frac{1}{2}s_H
c_H
   & \frac{1}{4}(2 s_H^2-1) 
 &  \frac{1}{4}(2 s_H^2-1) 
   \\
 0& - \frac{1}{2} s_H c_H&
\frac{1}{2}s_H c_H&\frac{1}{4}(1-2 s_H^2)  &
\frac{1}{4}(1-2 s_H^2)
   \\
 0 & \frac{1}{4}(2 s_H^2-1)& \frac{1}{4}(1-2 s_H^2)
   & -\frac{1}{2}s_H c_H&
-\frac{1}{2}s_H c_H\\
 0 & \frac{1}{4}(2 s_H^2-1) & \frac{1}{4}(1-2 s_H^2)
   & - \frac{1}{2}s_H c_H&
   -\frac{1}{2}s_H c_H
\end{array}\right)}_{G_{ht\bar{t}}/y}
\left(\begin{array}{c}
t_R\\ u_R\\u_{1R}\\t_{4R}\\T_{4R}
\end{array} \right)_{H=\langle H\rangle}\hspace*{-1cm} +h.c.\;,\label{mod11}
\end{equation}
\begin{equation}
-\mathcal{L}_{hb\bar{b}}=y\;h\overline{ \left(\begin{array}{c}
                      b_L \\d_L \\ d_{1L}\\ d_{4L} 
                      \end{array}\right)}
\underbrace{\left(
\begin{array}{cccc}
 0 & 0 & 0 & 0 \\
 0 &\frac{1}{2}s_H c_H& -\frac{1}{2}
   s_H c_H & \frac{1}{2 \sqrt{2}}(1-2 s_H^2) \\
 0&-\frac{1}{2}
   s_H c_H&\frac{1}{2}s_H c_H& 
  \frac{1}{2 \sqrt{2}}(2 s_H^2-1) \\
 0 & \frac{1}{2 \sqrt{2}}(1-2 s_H^2) & \frac{
  1}{2 \sqrt{2}}(2 s_H^2-1) & -s_H c_H
\end{array}
\right)}_{G_{h b\bar{b}}/y}
\left(\begin{array}{c}
                      b_R \\d_R \\ d_{1R}\\ d_{4R} 
                      \end{array}\right)_{H=\langle
H\rangle}\hspace*{-1cm}+h.c.\;.\label{mod12}
\end{equation}\normalsize
The matrices for the couplings to top-like states, $G_{ht\bar{t}}$,
and to bottom-like states, $G_{hb\bar{b}}$, in the mass eigenstate
basis are obtained by multiplication with the matrices $U_{L,R}$
defined in Eq.~\eqref{mod09}. The charge-5/3 fermions only interact with
the Higgs boson through small off-diagonal terms and are not relevant
for our analysis. Their coupling matrix is therefore not given
explicitly here. 
The couplings of the fermions to the gauge bosons are obtained from
Eq.~\eqref{mod08} in the interaction basis with subsequent rotation to
the mass eigenstates. In the following section also the couplings of the
fermions to the Goldstone bosons will be needed. They can be derived
from Eq.~\eqref{lagragian} by using Eq.~\eqref{mod10} and doing
the following replacements,
\begin{equation}
 h_1\to \frac{G^--G^+}{i\sqrt{2}}, \hspace*{1cm}  h_2\to-
\frac{G^-+G^+}{\sqrt{2}},\hspace*{1cm}  h_3\to G_0\;.\label{Goldstoneid}
\end{equation}
The couplings of the Goldstone bosons with the fermions can be found
in Appendix \ref{GOLDSTONECOUPLINGS}.

\section{Computation of electroweak precision
  observables \label{sec:ewpt}}
The results obtained at LEP put important constraints on New Physics
models. The data indirectly constrains physics at high energies which
enters in loop corrections to the observables at the electroweak
scale. In this section the contributions to the Peskin-Takeuchi $S$ 
and $T$ parameters~\cite{Peskin:1991sw} will be shortly
reviewed. Subsequently, the computation of the one-loop contributions to
the non-oblique corrections to the $Z b_L \bar{b}_L$ vertex due to the partial
compositeness of the bottom quark will be presented. The $U$ parameter
will not be discussed here, as it only receives contributions from operators of
dimension eight or higher. For convenience, we use instead of $S$, $T$
and the shift in the $Z b_L \bar{b}_L$ coupling the parameters
$\epsilon_1,\epsilon_2, \epsilon_3$ and
$\epsilon_b$~\cite{Altarelli:1990zd}, as they do not depend on a
reference point in the SM. 

\subsection{Contributions to $\epsilon_1$ \label{subsec:eps1}}
The $T$ parameter -- or equivalently $\epsilon_1$ -- gets 
a correction due to modified Higgs-vector boson couplings. They
prevent a full cancellation of the
UV-divergencies in the $T$-parameter so that a logarithmically
divergent part remains \cite{Barbieri:2007bh}. 
It is cut off by the mass of the first vector resonance $m_{\rho}$,
\begin{equation}
\Delta\epsilon_1^{IR}=-\frac{3\alpha(m_Z^2)}{16 \pi
\cos^2\theta_W}\xi\log\left(\frac{m_{\rho}^2}{m_h^2}\right)\;,
\end{equation}
with $\xi = v^2 / f^2$, {\it cf.}~Eq.~(\ref{eq:xidef}) and $\alpha$ the
electromagnetic coupling at the scale $m_Z$. The Weinberg angle is
denoted by $\theta_W$. Another
important contribution to $\epsilon_1$ comes from loops of fermionic
partners. Explicit formulae at the one-loop order can be found in
Refs.\cite{Lavoura:1992np,Anastasiou:2009rv}. 

\subsection{Contributions to $\epsilon_3$ \label{subsec:eps3}}
Similar to the IR contribution to $\epsilon_1$, a UV-divergent
contribution due to modified Higgs-vector boson couplings also arises for the $S$ parameter  -- or $\epsilon_3$ --,
\begin{equation}
\Delta\epsilon_3^{IR}=\frac{\alpha(m_Z^2)}{48 \pi
\sin^2\theta_W}\xi\log\left(\frac{m_{\rho}^2}{m_h^2}\right)\;.
\end{equation}
Additionally, at tree-level there is a UV contribution from the mixing of
elementary gauge fields with new vector and axial vector 
resonances \cite{Giudice:2007fh,Contino:2010rs},
\begin{equation}
\Delta\epsilon_3^{UV}=\frac{m_W^2}{m_{\rho}^2}\left(1+\frac{m_{\rho}^2}{m_a^2}
\right)\;,
\end{equation}
where $m_a$ denotes the mass of the first axial vector resonance. For
definiteness, we set $m_{\rho}/m_a\approx 3/5$ as obtained in the
five-dimensional $SO(5)/SO(4)$ models of 
Refs.~\cite{Agashe:2004rs,Contino:2006qr}. We explicitly checked that
varying $m_{\rho}/m_a$ between 1 and 2 has only a slight effect on our
numerical results. The finite fermionic one-loop contributions to
$\epsilon_3$, which can be found in Ref.~\cite{Lavoura:1992np}, are
neglected, as they are small compared to the tree-level UV
contributions given in~\cite{Anastasiou:2009rv}. As recently pointed out in
Ref.~\cite{Grojean:2013qca}, however, there can be an additional
logarithmically divergent contribution stemming from fermion loops,
which is given by 
\begin{equation}
 \Delta\epsilon_3^{div}\sim
\text{Tr}\left[W_L^{\dagger}Y_L+W_R^{\dagger}Y_R\right]\;,\label{eps3div}
\end{equation}
where $W_{L,R}$ are the left- and right-handed fermion couplings to
$W^{3}_{\mu}$ and $Y_{L,R}$ the corresponding hypercharges. In our case the
trace in Eq.~\eqref{eps3div} is zero. 

\subsection{Contributions to $\epsilon_b$ \label{subsec:epsb}}
Since light quarks are assumed to couple to any New Physics in a subdominant
way, no vertex corrections to the $e^+e^-$ annihilation process at LEP have to
be taken into account. The only exception is the $Z b_L \bar{b}_L$ vertex,
because the left-handed $b$-quark is in the same $SU(2)_L$ doublet as 
the top quark, which itself has a large mixing with composite
fermions. For this vertex, New Physics contributions can thus be sizeable.
\s

\begin{figure}[t!]
\vspace*{-3cm}
\begin{center}\hspace*{-3.5cm}
 \includegraphics[scale=1.0]{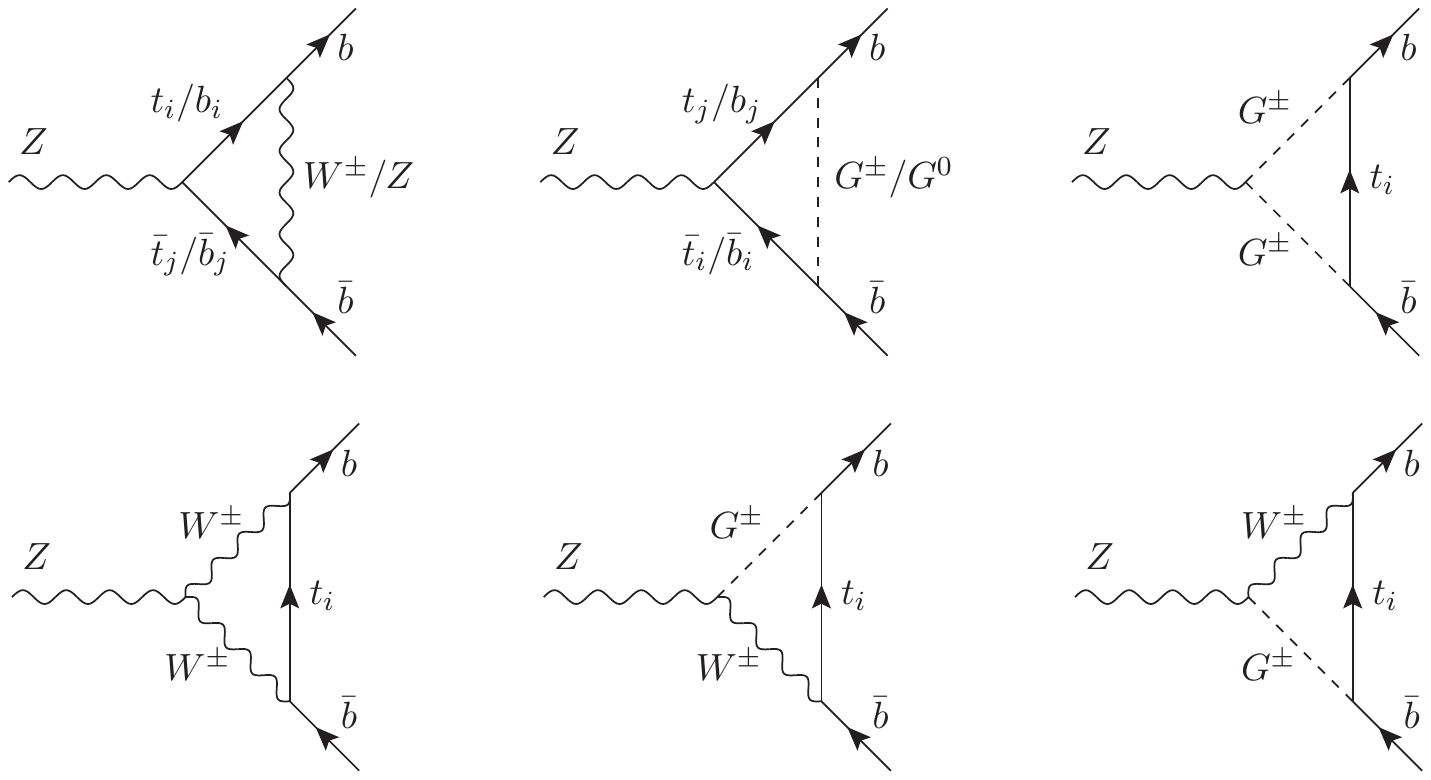}
\end{center}
\vspace*{-18cm}
\caption{Loop vertex diagrams for heavy fermion partner contributions to
  the $Z b_L\bar{b}_L$ coupling.}
\label{fig:001}
\end{figure}

The Lagrangian for the coupling of a $Z$ boson to a quark $\Psi_i^{Q}$
of charge $Q$ in the mass eigenstate basis is parameterized by
\begin{equation}
 \mathcal{L}_{Z}=\frac{g}{2 c_W}Z_{\mu}\bar{\Psi}_{Q}^{i}
\gamma^{\mu}\left(X_{ij}^{QL}P_L+X_{ij}^{QR} P_R-2 s_{W}^2
Q\delta_{ij}\right)\Psi_{Q}^j\;,\label{epsb01} 
\end{equation}
where $i,j$ run over all quarks present in the model. Here and below
we use the short-hand notation $c_W \equiv \cos \theta_W$ and $s_W
\equiv \sin \theta_W$. We keep the coupling
general so that the result can also
be applied to other cases. The decay amplitude of the $Z$ boson into a
pair of massless left-handed $b$-quarks is given by
\begin{equation}
 \mathcal{M}_{Z\to b_{\scriptscriptstyle L}\bar{b}_{\scriptscriptstyle
L}}=-\frac{e(g_{b_L}^{SM}+\delta g_{b_L})}{c_W
s_W}\epsilon_{\mu}(p_Z)\bar{b}(p_{\bar{b}})\gamma^{\mu}\frac{1-\gamma_5}{2}b(p_{
b } )\;,
\end{equation}
with the electric charge $e$ and the SM coupling $g_{b_L}^{SM}$ of the
$Z$ boson to the left-handed $b$-quarks. The polarization vector of the
$Z$ boson with four-momentum $p_Z$ is denoted by $\epsilon_\mu$. 
A left-right symmetry prevents $\delta g_{b_L}$, which contains the
effects from New Physics, from getting
tree-level contributions \cite{Agashe:2006at}. However, important 
contributions to $\delta g_{b_L}$ can occur through loops of new fermions. In
Fig.~\ref{fig:001} the Feynman diagrams for the one-loop corrections to
$Zb_L\bar{b}_L$ including gauge bosons and Goldstone bosons are shown. 
There are also diagrams involving the Higgs boson, which, however, have a
negligibly small contribution. 
In order to quantify the beyond the SM effect of the heavy quarks on
$\delta g_{b_L}$, the SM 
contribution $\mathcal{M}^{t+b}_{SM}$ of the bottom and top quarks has to be
subtracted,
\begin{equation}
 \delta g_{b_L}=\mathcal{M}^{heavy}-\mathcal{M}^{t+b}_{SM}\;,\label{epsb02}
\end{equation}
where $\mathcal{M}^{heavy}$ denotes the contributions from the loops
with the heavy quarks, the top and the bottom quark. The $Zb_L\bar{b}_L$ vertex
needs to be
renormalized
to become finite.  
We adopt an on-shell renormalization scheme similar to
Ref.~\cite{Denner:1991kt}. The wave function renormalization constants
$\delta Z^{L,R}$ relate the left- and right-handed bare fields $b_0^{L,R}$ with the
renormalized ones $b^{L,R}$,
\begin{equation}
 b_0^{L,R}=(1+\frac{1}{2}\delta Z^{L,R})b^{L,R}\;.
\end{equation}
The $Z$ boson coupling to the left-handed bottom-type quarks is
proportional to, {\it cf.}~Eq.~(\ref{epsb01}), 
\begin{equation}
X^{-1/3,L}=U_L^{b\;\dagger} T_{L}^3 U_L^{b}\;, \label{epsb03}
\end{equation}
where $T_{L}^3$ is the generator defined in Eq.~\eqref{mod01}.\footnote{For
the renormalization procedure the concrete definition of the 
generator $T_{L}^3$ does not matter, however. Our results are also 
applicable to other groups and hence different generators.}
For the renormalization of the mixing matrix $U^b_L$ in
Eq.~\eqref{epsb03} a counterterm $\delta u_L^b$ is introduced. The
complete $Z b_L\bar{b}_L$ vertex including the counterterm in the mass
eigenstate basis then reads
\begin{equation}
\begin{split}
 \mathcal{L}_{Z \bar{b}_L b_L}
=-&\frac{e }{s_{W} c_{W}} \bar{b}\,\gamma_{\mu} \left(1+\frac12 \delta
Z_{L}^{\dagger}\right) \left(1+\delta u_L^b\right)  U_L^b
\left(T_{L}^3-2 s_{W}^2 Q\right)\\&
U_L^{b\;\dagger}\left(1+\delta u_L^{b\;\dagger} \right)
\left(1+\frac12 \delta Z_{L}\right) P_L b\, Z^{\mu} \;,
  \end{split}
\end{equation}
where $P_L = (1-\gamma_5)/2$ denotes the left-handed projector. 
Note that only the wave function renormalization constants for the
$b$-quark fields and the counterterm of the mixing matrix are needed, 
whereas the electric charge, the Weinberg angle
and the wave function renormalization of the $Z$ boson are already
included in the oblique parameters \cite{Peskin:1991sw,
  Gonzalez:2011he}, due to their universality for all $Zf \bar{f}$
vertices. The counterterm is defined antihermitian, as the bare and
the renormalized mixing matrices are unitary, {\it
  cf.}~Ref.~\cite{Denner:1990yz},\footnote{The question of gauge
  invariance for this definition of the mixing matrix was widely
  discussed in the literature \cite{Gambino:1998ec, Yamada:2001px}. We follow
Ref.~\cite{Yamada:2001px} in order to obtain a gauge independent result. }
\begin{equation}
 \delta u_{L,ij}^b=\frac{1}{4} \left(\delta Z_{ij}^L-\delta
Z_{ij}^{L\;\dagger}\right)\;.
\end{equation}
Defining the structure ($P_R = (1+\gamma_5)/2$)
\begin{equation}
 \Sigma_{ij}(p^2)=\slashed{p}\Sigma_{ij}^L(p^2)P_L+\slashed{p}\Sigma_{ij}
^R(p^2)P_R+\Sigma_{ij}^l(p^2)P_L+\Sigma_{ij}^r(p^2)P_R
\end{equation}
for the fermion self-energy $\Sigma$, the wave function renormalization
constant $\delta Z^L$ for the left-handed fermion field is given by
\begin{align}
 \delta Z^L_{ij}&=\frac{2}{m_i^2-m_j^2}\widetilde{\text{Re}}\left(m_j^2
\Sigma^L_{ij}(m_j^2)+m_i m_j \Sigma^R_{ij}(m_j^2)+m_i\Sigma^l_{ij}(m_j^2)+m_j
\Sigma_{ij}^r(m_j^2)\right) &i\neq j\label{epsb04}\\
\delta Z^L_{ii}&=-\widetilde{\text{Re}}\,\Sigma_{ii}^L(m_i^2)-m_{i}
\frac{\partial}{ \partial
p^2}\widetilde{\text{Re}}\left(m_i(\Sigma^L_{ii}(p^2)+\Sigma^R_{ii}
(p^2))+\Sigma^l_{ii}(p^2)+\Sigma^r_{ii}(p^2)\right)\lvert_{p^2=m_i^2}  
&i=j \;,\label{epsb05}
\end{align}
where $\widetilde{\text{Re}}$ only takes the real part of the one-loop
integrals but keeps the complex structure of the parameters. 
Note that in our calculation we set the bottom mass to zero, which
implies that either $m_i$ or $m_j$ is zero in Eq.~\eqref{epsb04}
and that $m_i=0$ in Eq.~\eqref{epsb05}.  
\begin{figure}[t]
\vspace*{-2cm}
\begin{center}\hspace*{-0.5cm}
\includegraphics[scale=0.8]{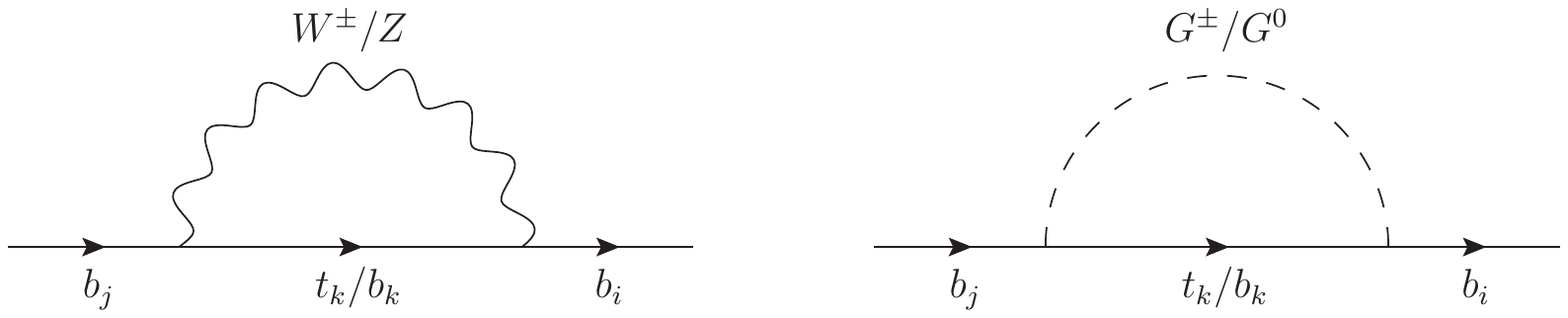}
\end{center}
\vspace*{-19cm}
\caption{Self-energy corrections needed for the renormalization of the
  vertex $Z b_L\bar{b}_L$.}
\label{fig:002}
\end{figure}
The Feynman diagrams of the self-energies which we need for the
renormalization of the $Zb_L \bar{b}_L$ vertex are shown in Fig.~\ref{fig:002}.
For the computation the programs {\tt FeynCalc}\cite{Mertig:1990an} and
{\tt FeynArts/FormCalc}\cite{Kublbeck:1990xc,Hahn:1998yk} were
used. The final result can be found in Appendix \ref{Appendixzbb}. It
is given in terms of general coupling factors so that it can be
applied to other cases. The notation is similar to the one
used in Ref.~\cite{Anastasiou:2009rv} so that the results can easily
be compared. The results obtained for the vertex diagrams in
Fig.~\ref{fig:001} agree with those of Ref.~\cite{Anastasiou:2009rv}.
The differences with respect to Ref.~\cite{Anastasiou:2009rv} arise
from the renormalization of the mixing matrix, which we performed and
which was not necessary in Ref.~\cite{Anastasiou:2009rv} as the
authors did not take into account the case of a bottom quark mixing
with heavy fermion partners. In our case, a finite result for $\delta
g_{b_L}$ can only be obtained if the renormalization of the
mixing matrix is included. \s

A comment is in order about contributions from the UV dynamics of the
theory to the EWPTs. In Ref.~\cite{Grojean:2013qca} it was shown that
there can be possibly large contributions to the $S$ parameter and the
$Zb_L\bar{b}_L$ coupling from a non-decoupling of UV-physics. This can
even give rise to logarithmically divergent contributions, as {\it e.g.} in the
$Zb_L\bar{b}_L$ coupling due to an effective 4-fermion operator. The coupling constant of this operator is not relevant for the rest of our
analysis and we therefore assume it to be small.
There could be further finite contributions from the UV dynamics of the
theory~\cite{Grojean:2013qca}, which we neglect, however, since there is
no reasonable way to estimate them in terms of the fields entering our
effective Lagrangian.

\subsection{The $\chi^2$ test and numerical results}
\label{chapnumewpt}
The agreement of our model with the experimental data can be assessed
by performing a $\chi^2$ test. 
The experimental values for the $\epsilon$ parameters and their
correlation $\rho$ come from the LEP measurement at the $Z$ pole mass, see
Ref.~\cite{ALEPH:2005ab}. We use, however, the updated values of
Ref.~\cite{Gillioz:2012se}, which take into account a newer value of the $W$
mass \cite{Aaltonen:2012bp}:
\begin{equation}
	\begin{array}{lcl}
\epsilon_1^{exp} & = & \left( 5.4 \pm 1.0 \right) \cdot
10^{-3}, \\
\epsilon_2^{exp} & = & \left( -7.9 \pm 0.90 \right) \cdot
10^{-3}, \\
\epsilon_3^{exp} & = & \left( 5.34 \pm 0.94 \right) \cdot
10^{-3}, \\
\epsilon_b^{exp} & = & \left( -5.0 \pm 1.6 \right) \cdot
10^{-3}, \\
	\end{array}
	\hspace{1.0cm}
	\rho = \left( \begin{array}{cccc}
		1 & 0.80 & 0.86 &  0.00 \\
		0.80 & 1 & 0.53 & -0.01 \\
		0.86 & 0.53 & 1 & 0.02 \\
		0.00 & -0.01 & 0.02 & 1
	\end{array} \right)\;.
\end{equation}
The theory contributions to the parameters $\epsilon_1,
\epsilon_2,\epsilon_3$ and $\epsilon_b$ are given by
\cite{Agashe:2005dk, Gillioz:2012se},
\begin{eqnarray}
 \epsilon_1^{th}&=&(5.66-0.86\log(m_h/m_Z))\cdot 10^{-3}+\Delta
\epsilon_1^{IR}+\alpha\Delta T\;,\nonumber\\
 \epsilon_2^{th}&=&(-7.11+0.16\log(m_h/m_Z))\cdot 10^{-3}\;,\nonumber\\
 \epsilon_3^{th}&=&(5.25+0.54\log(m_h/m_Z))\cdot
10^{-3}+\Delta\epsilon_3^{IR}+\Delta\epsilon_3^{UV}\;,\nonumber\\
 \epsilon_b^{th}&=&-6.48\cdot 10^{-3}+\delta g_{b_L}\;.
\end{eqnarray}
The first summands, respectively, are the SM corrections. The
contributions $\Delta 
\epsilon_i^{UV/IR}$ and $\delta g_{b_L}$ have been given 
in subsections \ref{subsec:eps1}\,--\,\ref{subsec:epsb}, and $\Delta T$
is the contribution to the $T$ parameter stemming from loops of heavy
fermions. The covariance matrix is defined by
\begin{equation}
 C_{ij}=\Delta \epsilon_i^{exp}\rho_{ij}\Delta \epsilon_j^{exp}\;,
\end{equation}
where $i,j$ runs over $1, 2, 3$ and $b$.
The parameters $\lambda_t$ and $\lambda_b$
are fixed by the requirement to recover the measured values of the top
and bottom quark masses, $\lambda_q$ has been traded for $\sin\phi_L$,
{\it cf.}~Eq.~(\ref{PCrotations}), and the scale $f$ is given by $f =
\sqrt{\xi} \, v$, so that the relevant set of free parameters for our
model is $\{ \xi, M_{10},\sin\phi_L, y, m_\rho \}$. The $\chi^2$ is
hence defined as
\begin{equation}
 \chi^2(\xi, M_{10}, \sin\phi_L, y,
m_{\rho})=\sum_{i,j}
 \left(\epsilon^{th}_i-\epsilon^{exp}_i\right)C^{-1}_{ij}
 \left(\epsilon^{th}_j-\epsilon^{exp}_j\right)\;.
 \label{eq:chi2EWPT}
\end{equation}
The electroweak precision data indicate a preference for a heavy
vector resonance, so that we fix the parameter $m_\rho$ to its maximal
value of $4\pi f$ required by perturbativity. We found that this leads 
for most of the parameter sets to minimal or  close to minimal values
of $\chi^2$. We are therefore left with four degrees of freedom $\{\xi, M_{10},
\sin\phi_L, y\}$. A specific point in the parameter space fulfills the
electroweak precision tests at 99\% C.L.~if it satisfies the criterion 
\begin{equation}
 \chi^2(\xi, M_{10}, \sin\phi_L, y) - \chi_{min}^2\leq 13.28\;,
\end{equation}
where $\chi_{min}^2$ is the minimum of $\chi^2$ with
$\chi_{min}^2\approx 0.87$. This is smaller than the SM value 
$\chi^2\approx 4.71$ as expected for a model with additional
parameters. \s

A further constraint on the model is imposed by the recent measurement
of the single top cross section at CMS\cite{CMSsingletop}, providing a
lower limit on the CKM matrix element of $\lvert V_{tb}\rvert> 0.92$
at 95\% C.L.. The constraint on $V_{tb}$ will be discussed in more detail in
Section~\ref{sec:numerical}. \s
 
We performed a scan over the parameter space, setting the top and
bottom quark masses to $m_t=173.2$ GeV and $m_b= 4.2$~GeV,
respectively, and the Higgs boson mass to $m_h=125$ GeV. For the
vector bosons masses we used $m_W=80.385$ GeV and
$m_Z=91.1876$~GeV. The model parameters have been varied in the range 
\begin{equation}
 0 \leq\xi\leq 1\;,\hspace*{1cm} 0<\sin\phi_L\leq 1\;,\hspace*{1cm}
 |y|<4\pi\;, 
\hspace*{1cm} 0\leq M_{10}\leq 10 \text{ TeV}\;.\label{pararange}
 \end{equation}

\begin{figure}\hspace{-0.1cm}
   \includegraphics[scale=0.31, angle=-90]{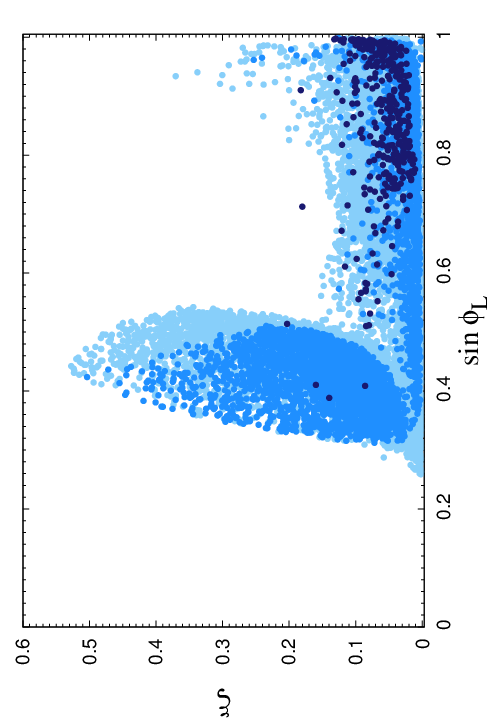}\hspace*{0.5cm}
    \includegraphics[scale=0.30, angle=-90]{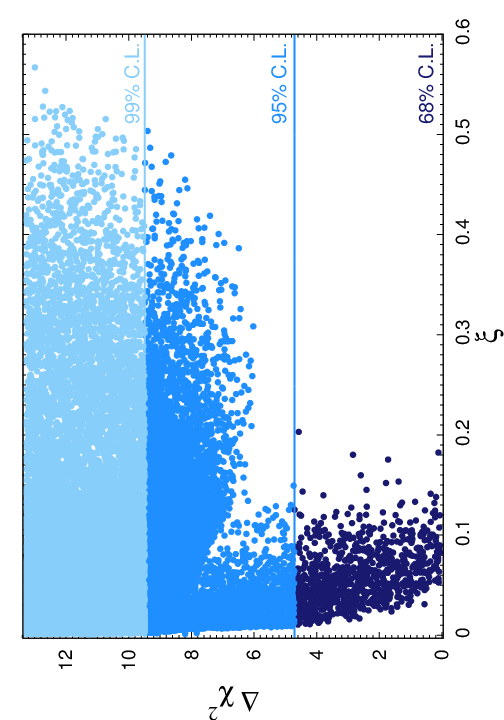}
\caption{Parameters passing the $\chi^2$ test of electroweak precision
  observables, fulfilling in addition $|V_{tb}| > 0.92$, for a scan
  over $\xi$, $\sin\phi_L$, $y$ and $M_{10}$. Dark blue: $68\%$
  C.L. region, medium blue: $95\%$ C.L. region and light blue: $99\%$
  C.L. region. {\it Left:} the $\xi$-$\sin\phi_L$ plane. {\it Right:}
  $\Delta\chi^2$ versus $\xi$.}
\label{fig:003}
  \end{figure}

In addition we only retained points for which $|V_{tb}| > 0.92$. 
The result of the scan is shown in Fig.~\ref{fig:003}~(left) in the
$\xi$-$\sin\phi_L$ plane. As can be inferred from the plot, for
non-vanishing left-handed compositeness of the top and bottom quark,
values of $\xi$ close to 0.2 are still allowed at 68\% C.L.. For
intermediate values, $0.4 \lsim \sin\phi_L \lsim
0.5$, parameter points with $\xi$ as large as $\xi \sim 0.5$ pass the
constraints.\footnote{In
  Ref.~\cite{Gillioz:2008hs} a similar plot as the one of
  Fig.~\ref{fig:003} (left) was shown for the fundamental
  representation, and a maximal allowed $\xi$ value of only
  $\xi_{max}\approx 0.35$ was found. We use a different
  representation for the extra fermion multiplet, however. Furthermore, instead
  of $m_{\rho}=2.5$~TeV in \cite{Gillioz:2008hs} we take $m_{\rho}=4
  \pi f$ which lowers the tension with the electroweak precision observables.}
In case of mostly composite left-handed quarks, $\sin \phi_L \gsim 0.9$, 
the constraints are passed at 99\% C.L. for $\xi$ values up to about 0.35.
It is the positive fermionic contributions to the $T$ parameter which
drive it back into the region compatible 
with EWPTs.\footnote{For a
comprehensive discussion (in the fundamental representation), see
Ref.~\cite{Gillioz:2008hs}.} 
For $\sin\phi_L\lsim 0.25$, there are
no allowed points, as for too low values of $\sin\phi_L$ the
correct top mass cannot be obtained, {\it cf.}~Eq.~\eqref{mod13}.
The bottom quark being in the same doublet as the top quark, is hence
mostly left-handed composite, as $\sin\phi_{Rb}$ must be small enough in
order not to generate a too large bottom mass, {\it cf.}~Eq.~(\ref{mod13}).
Figure~\ref{fig:003} (right) shows $\Delta\chi^2 \equiv \chi^2 -
\chi^2_{min}$ versus $\xi$. The smallest values for $\Delta\chi^2$
are obtained for $0.01<\xi<0.2$. In contrast, high values of $\xi$
lead to large $\Delta \chi^2$, corresponding to a compatibility with the EWPT at
99\% C.L.. Note that the SM limit is  
obtained for $\xi\to 0$ and $M_{10}\to\infty$. Due to the restriction
of the scan to $M_{10}\leq 10$~TeV, it is not contained in the plot. \s

The impact of the bottom quark and its partners on the $\chi^2$ test
is significant. Their inclusion not only requires the renormalization
of the mixing matrix, which influences
the finite terms. For some parameters in our scan the inclusion
of the bottom partners in the loops can also change $\Delta \chi^2$ by
a factor of 2.  For the majority of the parameter points, however, the
effect is much smaller.  The contributions from diagrams with Higgs
bosons in the loops alter $\Delta \chi^2$ by $\pm 2.9$\% at most, for
most parameter sets even less. \s

A comment is in order about the approximation of zero bottom quark
mass in the computation of the corrections to the $Zb_L \bar{b}_L$
vertex. Neglecting the bottom mass changes the couplings of the
bottom quark and of the bottom-like quarks to the vector bosons and
Goldstone bosons. The effect, however, is small. The matrix
element $X^{-1/3,L}_{bb}$, {\it cf.}~Eq.~(\ref{epsb03}), changes by
maximally $1\%$ and the change in the corresponding matrix element for the
Goldstone coupling is $\mathcal{O}(m_b/v)$. Compared to the largest
matrix elements in the Goldstone coupling matrix this is less than a
percent effect.\footnote{We discuss here the Goldstone
  coupling as this would correspond to the gauge-less limit in which
  {\it e.g.} in Ref.~\cite{Gillioz:2008hs} the EWPT were obtained for
  the fundamental representation.} We explicitly verified this numerically.
Additional mass terms can arise in the loop corrections to the
$Zb_L\bar{b}_L$ vertex. They are proportional to $m_b/m_Z$, and
assuming that the couplings multiplying these terms are of
the same order as the ones multiplying $m_t/m_Z$, they would only
contribute to about 3\% of the total matrix element. A conservative estimate of
the error done by neglecting the bottom mass is therefore $5\%$, obtained by
adding up linearly the error due to the kinematics and an estimate of $2\%$ for
the error due to the change in the couplings. \s

\begin{figure}\hspace{-0.1cm}
\begin{center}
   \includegraphics[scale=0.31, angle=-90]{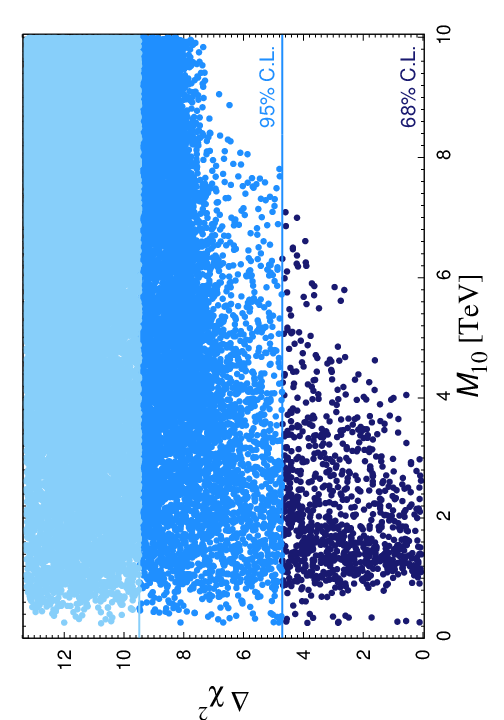}
\caption{\label{fig:ewptmass10}$\Delta\chi^2$ versus $M_{10}$ of the
  parameters passing the 
  $\chi^2$ test of electroweak precision 
  observables, fulfilling in addition $|V_{tb}| > 0.92$, for a scan
  over $\xi$, $\sin\phi_L$, $y$ and $M_{10}$. Dark blue: $68\%$
  C.L. region, medium blue: $95\%$ C.L. region and light blue: $99\%$
  C.L. region.}
\end{center}
  \end{figure}

As mentioned earlier loop contributions to the $T$ parameter from the
top and bottom partners are important to render the model compatible
with the EWPT for non-vanishing $\xi$ values. The implications of the
electroweak precision data on the masses of the vector-like quarks can
be inferred from Fig.~\ref{fig:ewptmass10}. It shows 
$\Delta\chi^2$ as a function of $M_{10}$ which sets the scale for the
top and bottom partner masses. As
expected, the best compatibility of the model with the electroweak
precision observables is obtained for non-vanishing masses of the
order of 200~GeV$\lsim M_{10} \lsim$5~TeV.  The bulk of the masses for
the points which are best compatible with EWPT lies between about 800
GeV and 1.6 TeV, however. This is compatible with the lower limits
from direct searches for heavy fermions, as will be discussed in
detail in the next section.  

\section{Constraints from Higgs Results and Direct Searches for Heavy Fermions \label{sec:higgsres}}
Further constraints on Composite Higgs Models come from the LHC Higgs
search results. Both production processes and decay rates of the Higgs
bosons are modified compared to the SM \cite{Espinosa:2010vn}. The
modifications arising in our model shall be presented in the
following. Subsequently, the constraints due to direct LHC searches
for heavy fermions will be discussed.
\subsection{Higgs Boson Production
Processes}
{\bf \underline {Gluon fusion:}} Gluon fusion \cite{georgi} is
the main Higgs production mechanism at the LHC and mediated already at
leading order by loops of heavy quarks. In addition to the top and
bottom quark loops present in the SM, in Composite Higgs Models also
heavy quark partners contribute and the Higgs Yukawa couplings are
modified.\footnote{For a general discussion of the effects of
  additional heavy quarks on (multiple) Higgs production through gluon
  fusion, taking into account experimental bounds, see 
  Refs.~\cite{Dawson:2012di,Dawson:2012mk}.} The QCD corrections to
the process 
are important. In the SM they have been obtained at next-to-leading 
order (NLO) including the full quark mass dependence and in the heavy
top mass limit \cite{ggnlo}. They increase the cross section by
50-100\%. At next-to-next-to-leading order (NNLO) QCD they are known in 
the heavy top quark limit \cite{ggnnlo}, adding another 20\%. Top
quark mass effects on the NNLO cross section have been investigated in
Ref.~\cite{topmasseffect}. A resummation of soft gluons has been
performed at  next-to-next-to-leading log (NNLL)
accuracy~\cite{resum}. First results for the $\text{N}^3$LO QCD
corrections have been given in Refs.~\cite{ggnnnlo}.  
For Composite Higgs Models the QCD corrections up to NNLO were
calculated in  Ref.~\cite{Furlan:2011uq}, keeping the full bottom mass
dependence through NLO. The two-loop Yukawa
corrections to gluon fusion in the top partner singlet model have been
presented in \cite{Dawson:2013uqa}. Note, that in Composite Higgs
models without new heavy fermion partners the QCD corrected SM cross
section can be taken over by adjusting the
Higgs-Yukawa couplings. This cannot not be done, however, for the
electroweak corrected process \cite{ewcorr}.  \s 

We implemented our model in the Fortran code {\tt HIGLU} \cite{higlu}
in order to obtain the NLO QCD corrections with full mass dependence
on the quark masses. This was done similar to the implementation of
the 4th generation in {\tt HIGLU} \cite{higluweb}. The Higgs Yukawa couplings
had to be adjusted and all summations were extended to also include 
the loops with the new fermions. Electroweak corrections
in Composite Higgs Models are not available and NNLO QCD 
corrections are only available in the heavy top quark limit, which
cannot be applied for the bottom quark. We therefore only take into
account the NLO QCD corrections. The $K$-factor obtained in this way, 
\begin{equation}
 K=\frac{\sigma_{NLO}}{\sigma_{LO}} \, ,
\end{equation}
is roughly the same as in the SM for NLO QCD corrections, up to deviations of
less then $2\%$ depending on the specific parameter point, in agreement with
Ref.~\cite{Furlan:2011uq}. \s

In
Ref.~\cite{Falkowski:2007hz,Low:2010mr,Azatov:2011qy,Gillioz:2012se}
it was shown by applying the low-energy theorem
\cite{Kniehl:1995tn} that the leading order gluon fusion cross
section $\sigma$ with fermions in the fundamental representation and
neglecting the mixing of the bottom quark with heavy partners, is
given by the pure Higgs non-linearities, 
\begin{equation}
 \frac{\sigma}{\sigma_{SM}}\approx \frac{(1-2\xi)^2}{(1-\xi)}\;,
\end{equation}
where $\sigma_{SM}$ denotes the SM gluon fusion cross section.
The cross section, which only depends on $\xi$ but not on the details of
the spectrum of the new fermions, is therefore always reduced compared to
the SM for $\xi<0.75$. This result does not hold any more, however, if
there exists a mixing with bottom partners \cite{Azatov:2011qy,
  Delaunay:2013iia}. For the bottom quark the LET
cannot be applied and the matrix element for the 
bottom-like contributions ${\cal M}_{bot}$ is given by
\begin{equation}
\mathcal{M}_{bot}\approx\mathcal{M}^{SM}_{LET}\left(\frac{1-2\xi}{\sqrt{1-\xi}}
-\frac{y_b}{y_{SM}} \right)\;,
\end{equation}
with ${\cal M}^{SM}_{LET}$ denoting the SM matrix element in the LET
approximation, and $y_b$ and $y_{SM}$ being the bottom quark Yukawa
coupling in our model and the SM, respectively. The gluon fusion cross
section thus depends on the details of the spectrum through
$y_b$. In Ref.~\cite{Azatov:2011qy} it was shown that this can even lead to
an enhancement of the cross section for the gluon fusion
process compared to the SM. 

\noindent
{\bf \underline{Vector boson fusion:}}
Vector boson fusion \cite{wzfusion} constitutes the next important
Higgs production mechanism after gluon fusion. In the SM, the NLO QCD
corrections to vector boson fusion are of $\mathcal{O}(10\%)$ of the
total cross section \cite{vbfnloqcd, Spira:1997dg}, the NNLO
QCD corrections are at the percent
level~\cite{Bolzoni:2010xr}. Electroweak corrections have been given
in \cite{Ciccolini:2007ec} and are of ${\cal O}(5\%)$. \s

In our model, the cross section at NLO QCD can be obtained from the
SM cross section by multiplying it with a factor $(1-\xi)$ stemming from the
modified Higgs couplings to massive vector bosons $V$ due to the Higgs
non-linearities, {\it cf.}~Eq.~(\ref{eq:higgsv}), 
\begin{equation*}
\sigma_{VBF}^{CHM}=(1-\xi)\sigma_{VBF}^{SM} 
 \;.
\end{equation*}
The cross section is reduced compared to the SM cross section, which
we calculated at NLO QCD with the Fortran code {\tt VV2H} \cite{V2HV}. 
Again, neither electroweak (EW) corrections nor NNLO QCD corrections
can be taken into account. \s

\noindent
{\bf \underline{Higgs-strahlung:}}
In Higgs-strahlung the Higgs boson is radiated off vector bosons. The
NLO QCD corrections increase the cross-section by ${\cal O} (30\%)$
\cite{Spira:1997dg,V2HVnlo}, the NNLO QCD corrections are small
\cite{Hamberg:1990np}. The electroweak corrections for the SM decrease
the cross section by $\mathcal{O}(5-10\%)$~\cite{Ciccolini:2003jy}.
We proceed analogously to vector boson fusion and only take into account NLO QCD
corrections. The SM cross section at NLO
QCD\cite{Spira:1997dg,V2HVnlo} has been computed with the code {\tt
  V2HV}\cite{V2HV} and subsequently multiplied with the appropriate
modification factor to obtain the Composite Higgs production cross section,
\begin{equation}
 \sigma_{Wh/Zh}^{CHM}=(1-\xi)\sigma_{Wh/Zh}^{SM} 
 \;.
\end{equation}

\vspace*{0.2cm}
\noindent
{\bf \underline{Associated production with top quarks:}}
The cross section for associated production of a SM Higgs boson of
$m_h=125$~GeV with a top quark pair \cite{lotth} is two orders of
magnitudes smaller than the gluon fusion cross section. We took the SM
cross section including NLO QCD corrections \cite{ttHnlo} from the LHC cross
section working group \cite{LHCcxn} and modified it to take into account
the Higgs-top Yukawa coupling of our model,
\begin{equation}
 \sigma^{CHM}(t\bar{t}h)=\left(\frac{g_{tth}}{g_{tth}^{SM}}\right)^2\sigma^{SM}
(t\bar { t } h)\;.
\end{equation}
The coupling $g_{tth}$ is obtained from the matrix Eq.~\eqref{mod11}
after rotation to the mass eigenstates.

\subsection{Higgs Boson Decays}
The Composite Higgs branching ratios have been calculated with the
Fortran code  {\tt HDECAY}\cite{hdecay}, which we have adapted to our
model\footnote{For a recent discussion on the implementation of the
  effective Lagrangian for a light Higgs-like boson into automatic
  tools for the calculation of Higgs decay rates,
  see~Ref.~\cite{Contino:2013kra}. The Fortran code {\tt eHDECAY} including
  the effective Lagrangian parametrization can be found at~\cite{ehdecay}.
  An implementation in {\tt FeynRules} has been provided in
 Ref.~\cite{Alloul:2013naa}. } by proceeding as follows:
To get the Composite Higgs fermionic decay widths, all
corresponding SM widths have been modified as
\begin{equation}
\Gamma^{CHM}_{h\to
f\bar{f}}=\begin{cases}
\left(\frac{(U_{L}^{b\dagger}G_{hb\bar{b}}U_{R}^b)_{b\bar{b}
}}{g_{hb\bar{b}}^{SM}}
\right)^2\Gamma^ {SM } _ { h\to f\bar{f}}\hspace*{1cm}&\mbox{if }f=b\;,\\
           \frac{(1-2 \xi)^2}{1-\xi}\Gamma^{
SM } _ { h\to f\bar{f}}\hspace*{1cm}&\mbox{if }f=c,s,\mu,\tau\;.
          \end{cases}
\end{equation}
The decays into top quarks are not relevant for a 125~GeV Higgs
boson. In the decay width into bottom quarks the factor
$(U_{L}^{b\dagger}G_{hb\bar{b}}U_{R}^b)_{b\bar{b}}$ denotes the matrix
element relevant for the bottom quark coupling after rotation of the
Higgs Yukawa coupling matrix $G_{hb\bar{b}}$, Eq.~(\ref{mod12}), 
into the basis of the mass eigenstates. The prefactor for the decays
into the charm ($c$), strange ($s$), muon ($\mu$) and $\tau$ final
states, which are elementary particles in contrast to the top and
bottom quark, is due to the Higgs non-linearities, implying a Yukawa coupling 
\beq 
g_{hf\bar{f}}^{CHM} = \frac{1-2\xi}{\sqrt{1-\xi}} \, g_{hf\bar{f}}^{SM}
\eeq
for the fermions in the fundamental and antisymmetric representation
\cite{Contino:2006qr}. 
The decays into vector bosons $V$ are obtained from the
corresponding SM widths by
\begin{equation}
\Gamma^{CHM}_{h\to VV}=(1-\xi)\, \Gamma^{SM}_{h\to VV}\;.
\end{equation} 
For the loop-induced decays also the top and bottom partners have to
be taken into account. The decay widths $h\to\gamma\gamma$ and $h\to
gg$ (at leading order) are modified as
\beq
\Gamma_{\gamma\gamma} =\frac{G_F \alpha^2 m_h^3}{128 \sqrt{2} \pi^3}
&&\hspace*{-0.5cm}\left| \,
\sum_{i=1}^{5}\frac{16}{9}\frac{v(U_L^{t\dagger}G_{htt}U_R^t)_{ii}}{m_{t_i}} \,
  A_{1/2}(\tau_{t_i}) 
+\sum_{i=1}^4 \frac{4}{9}\frac{v (U_L^{b\dagger}G_{hbb}U_R^b)_{ii} }{m_{b_i}}\,
A_{1/2}(\tau_{b_i}) \right.
\nonumber 
\\ 
&&\hspace*{-0.5cm}\left. \,+ \frac{4}{3} \frac{1-2\xi}{\sqrt{1-\xi}}
  \, A_{1/2} (\tau_\tau)
+\frac{16}{9}\frac{1-2\xi}{\sqrt{1-\xi}}\,A_{1/2}(\tau_c)
+\sqrt{1-\xi} A_{1}(\tau_W)\right|^2\;,\label{hgammadec}\\
\Gamma_{gg} =\frac{G_F \alpha_s^2 m_h^3}{36 \sqrt{2} \pi^3} 
&&\hspace*{-0.5cm}\left|\,
\sum_{i=1}^5 \frac{v(U_L^{t\dagger}G_{htt}U_R^t)_{ii}}{m_{t_i}} 
A_{1/2}(\tau_{t_i})
+\sum_{i=1}^4\frac{v(U_L^{b\dagger}G_{hbb}U_R^b)_{ii}}{m_{b_i}}
A_{1/2}(\tau_{b_i}) \right. \nonumber \\
&&\hspace*{-0.5cm}\left. \,+ \frac{1-2\xi}{\sqrt{1-\xi}} A_{1/2}
  (\tau_c) \right|^2\;, 
\eeq
where we introduced the notation
\begin{equation}
\tau_W=\frac{4 M_W^2}{m_h^2}
,\hspace*{0.5cm}\tau_{t_i/b_i}=\frac {4
  m_{t_i/b_i}^2}{m_h^2},\hspace*{0.5cm} \tau_c=\frac{4
m_c^2}{m_h^2}\hspace*{0.5cm}\text{and}\hspace*{0.5cm}\tau_{\tau}=\frac{4
m_{\tau}^2}{m_h^2}\;. 
\end{equation}
The masses of the top quark and its four heavy partners are denoted by
$m_{t_i}$ ($i=1,...,5$), the masses of the bottom quark and its three
heavy partners by $m_{b_i}$ ($i=1,...,4$), $m_c$ is the charm quark mass and
$m_{\tau}$ the mass of the $\tau$-lepton. The loop functions are given by
\begin{equation}
A_1(\tau)=-[2+3\tau+3\tau (2-\tau)f(\tau)]
\end{equation}
for $W$ bosons in the loop, and 
\begin{equation}
A_{1/2}(\tau)=\frac32 \tau [1+(1-\tau)f(\tau)] \; , 
\label{eq:048}
\end{equation}
for fermions in the loop, with
\begin{align}
f(\tau)&=\begin{cases}
\arcsin^2 \frac{1}{\sqrt{\tau}}& \text{for }\, \tau \geq 1 \; ,\\
-\frac14 \left[\log(\frac{1+\sqrt{1-\tau}}{1-\sqrt{1-\tau}})-i\pi \right]^2&
\text{for }\, \tau < 1 \; .
\end{cases}
\end{align}
Remark that in the LET the contribution due to the loops of the top
quark and its partners reduces to the pure Higgs non-linearities which
means that it is simply 
given by the SM top loop contribution modified with the coupling
factor $(1-2\xi)/\sqrt{1-\xi}$, parallel to the charm and $\tau$ loop
contributions. For the bottom loops, where the LET cannot be applied,
this is not the case. \s

We do not give an explicit formula for the decay $h\to Z\gamma$ as we
will not investigate this channel any further, which due to its
smallness practically does not affect the total decay
width.\footnote{A recent discussion on $h\to Z\gamma$ can be found in
  \cite{Azatov:2013ura}.} 
All decays are taken at NLO QCD if available in {\tt HDECAY},
see~\cite{Contino:2013kra,ehdecay} for details of the
implementation. Neither  
electroweak corrections nor NNLO QCD corrections were taken into
account. For slight deviations from the SM, EW corrections can be 
included as described in Ref.~\cite{Contino:2013kra}. We will
nevertheless neglect them as we also want to deal with possibly large
values of $\xi$. 

\subsection{Constraints from Searches for Heavy Fermions and from Flavour Physics}

The strongest bounds from direct searches for new vector-like fermions come
from ATLAS \cite{directsearchesATLAS1, directsearchesATLAS2} and CMS
\cite{directsearchesCMS, Chatrchyan:2013uxa}. Recently, both collaboration have
provided
direct bounds on the mass 
of the new fermions as a function of their branching ratios into SM
particles \cite{directsearchesATLAS1,directsearchesATLAS2,directsearchesCMS,
Chatrchyan:2013uxa},
since the
fermion pair production is a pure QCD process, which only depends 
on the mass of the particle, and can be computed independently of the model. 
The new top-like quarks can decay into $Wb$, $ht$ or $Zt$, the new
bottom-like fermions into $Wt$, $Zb$ or $hb$ and  the new charge-5/3
fermions into $Wt$. We have calculated the decay widths in our model
using the formulae of Ref.~\cite{Gillioz:2012se} (see also
Ref.~\cite{Bini:2011zb}), and directly compared them with the bounds
quoted by the collaborations. The bounds are obviously valid
for the lightest of the composite fermions, but not necessarily for the heavier
ones. The reason is that a composite fermion, which is massive enough
to decay into a lighter composite fermion and a $W$ or $Z$ boson,
could have a substantial decay width into the corresponding channel,
hence its branching ratios into the SM particles would be reduced. \s

In the specific model studied in this work, the situation is made
quite simple since the lightest of all composite fermions is always a
fermion of charge 5/3, decaying therefore 100\% into $Wt$. The
strongest bound on charge-5/3 fermions comes from the CMS
analysis~\cite{Bhattacharya:2013poa}, 
\begin{equation}
	m_\chi \geq 770~\textrm{GeV}.
	\label{eq:directdetectionbound}
\end{equation}
The bound on the bottom-like quarks turns out to be less stringent
than for the charge-5/3 fermions,\footnote{The search strategy for
  bottom-like quarks decaying mostly into $Wt$ is very similar to the
  search for a charge-5/3 fermion, since in both cases a final state
  is considered 
  with two same-sign leptons and a number of jets. However, in the
  case of the charge-5/3 fermions, the leptons come from the cascade
  decay $\chi \to W t \to WWb$ of a single  fermion with charge $\pm
  5/3$, while its antiparticle decays purely hadronically and its
  mass can be reconstructed from the jets, hence giving a stronger
  constraint than for a bottom-like quark.} 
but for top-like quarks ATLAS has limits extending up to around
850~GeV in the case of a decay mostly in $ht$~\cite{directsearchesATLAS1}. 
This limit can be applied as it is to the lightest of the charge-2/3
fermions, since it is in any case below the threshold for the decay of
a heavy top-like partner into $\chi W$ due to the bound of
Eq.~\eqref{eq:directdetectionbound}. 
In our model, however, the search for top-like fermions is never more
constraining than the search for the charge-5/3 ones.
In the future, and mostly with the LHC operating at 14~TeV,
important bounds will be derived from single production of a heavy vector-like
fermion, see {\it e.g.}~\cite{Aguilar-Saavedra:2013qpa,Andeen}, but such
bounds are not yet available.  \s

\begin{center}
\begin{figure}[t]
\centering
   \includegraphics[scale=0.31, angle=-90]{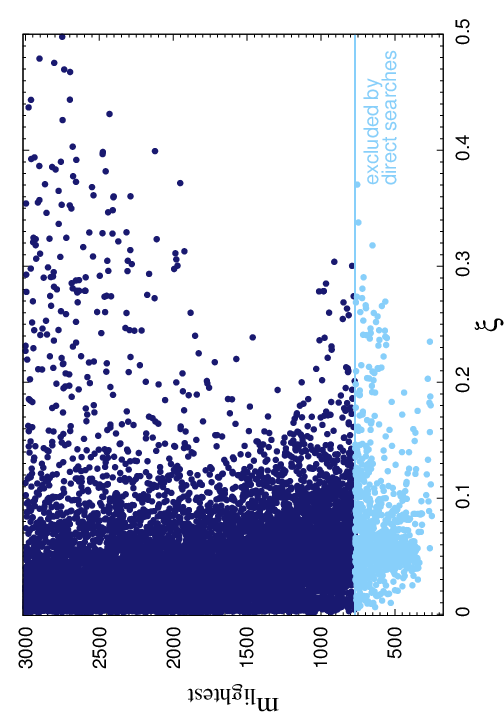}
        \caption{Spectrum of the lightest composite fermion as function of
$\xi$. The points in the plot are obtained from a scan over $\xi,\,y,\, M_{10}$
and $\sin\phi_L$ and fulfill the EWPT at 99\% C.L. and $|V_{tb}|>0.92$. The
light
blue points are excluded by direct searches for vector-like fermions
at 95\% C.L., the dark blue points are not excluded.}
\label{fig:spectrum}
  \end{figure}%
  \end{center}

In Fig.~\ref{fig:spectrum} we show the mass of the lightest composite
fermion as a function of $\xi$. The points in the plot are
the ones which pass the EWPT at 99\% C.L. and fulfill
$|V_{tb}|>0.92$. The light blue points are excluded by direct searches
at 95\% C.L., the dark blue points are not excluded. The
line in the plot marks the exclusion limit from CMS of 770 GeV on charge-5/3
fermions. As can be inferred from the plot this exclusion limit
eliminates quite some parameter space for $\text{m}_{\text{lightest}}
> 770$~GeV. No points are excluded above masses of the lightest
partner of 770 GeV which confirms that the bounds on heavy top 
partners of up to 850 GeV for large branching ratios of $T\to hb$ do not lead to
any additional constraints. \s

Flavour physics can lead to further constraints on Composite Higgs Models.
They depend, however, on the exact flavour structure of
the model. Anarchic flavour structures seem to be strongly constrained by CP
violating observables in the Kaon
system\cite{Csaki:2008zd}. Implementing minimal flavour violation can,
however, avoid these constraints \cite{Redi:2011zi}.
In this case, also the light quarks are required to be composite,
which can significantly change the Higgs phenomenology 
\cite{Delaunay:2013iia}. 
While dijet searches put constraints on the up and down quarks
\cite{dijetsearch}, the second generation quarks are practically not
constrained \cite{DaRold:2012sz}. Alternatively, the top quark can be treated
differently than the light quarks \cite{Redi:2012uj}. The
flavour bounds can still be satisfied, and the constraints from EWPT and
searches for compositeness are relaxed, as the first two generations are mostly
elementary. Both the left-handed and right-handed top can be composite
in this case. Bounds on the masses of the lightest fermionic resonance
have been obtained in Ref.~\cite{Barbieri:2012tu} and depend on the
specific flavour symmetry. 
We do not assume a specific flavour model and therefore do not further
discuss constraints from flavour physics. 
For additional discussions of flavour constraints on
Composite Higgs Models, see {\it e.g.}~Ref.~\cite{flavourconstraints}. 
%

\section{Numerical results \label{sec:numerical}}

In this section, we show numerical results for a combined analysis
taking into account the constraints from electroweak precision
observables, Higgs search results, the measurement of $V_{tb}$ and the
direct searches for heavy fermions. We make a random scan over the
parameter ranges defined in 
Eq.~\eqref{pararange} and with the SM input values as given in section 
\ref{chapnumewpt}. In order to test the agreement of our model with
the aforementioned constraints we perform a global $\chi^2$ test
similar to that of Refs.~\cite{Belyaev:2013ida}, 
\begin{equation}
 \chi^2= \chi^2_{EWPT}+\chi^2_{Higgs}
	+\chi^2_{V_{tb}}\;. \label{eq:chialltest}
\end{equation}
Notice that the constraints from direct searches of  new heavy
fermions are not included in the global $\chi^2$ test, but rather
imposed directly by only taking into account points which are not 
excluded at $95\%$ C.L. by direct searches. The $\chi^2_{EWPT}$ is the
$\chi^2$ for the electroweak precision tests defined in
Eq.~\eqref{eq:chi2EWPT}. \s
 
Regarding the constraints from the Higgs boson, the ATLAS and CMS
collaborations  provide the signal strengths
\begin{equation}
	\mu(X) = \frac{\sigma(pp \to h) \cdot BR(h \to X)}{\sigma(pp \to h)_{SM} \cdot BR(h \to X)_{SM}}
\end{equation}
including the correlations between the combination of the vector boson
fusion (VBF) and the Higgs-strahlung (VH) production modes (VBF+VH) and the
combination of gluon fusion (ggF) and the associated production
with a top quark pair (tth) (ggF+tth)~\cite{ATLASdata,
  CMSdata}. The results have been given as likelihood contours, which
correspond approximately for each Higgs boson decay channel to the ellipses
obtained from a $\chi^2$ test with two variables. We can therefore
write 
\begin{equation}
 \chi^2_{Higgs} = \sum_{\scriptscriptstyle
channels}\sum_{\scriptscriptstyle i,j=1,2}(\mu_i^{exp}-\mu_i^{th})C_{ij}^{-1
} (\mu_j^{exp}-\mu_j^{th})\;,
\end{equation}
where the best-fit points from the experiments are denoted by
$\mu^{exp}_1 = \mu^{exp}_{\scriptscriptstyle ggF+tth}$ and
$\mu^{exp}_2 = \mu^{exp}_{\scriptscriptstyle VBF+VH}$ and the
covariance matrix $C$ is defined as 
\begin{equation}
 C=\left(\begin{array}{c c}
       \Delta \mu_{\scriptscriptstyle ggF+tth}^2 & \rho \Delta
\mu_{\scriptscriptstyle ggF+tth}\,\Delta \mu_{\scriptscriptstyle VBF+VH}\\
        \rho \Delta \mu_{\scriptscriptstyle ggF+tth}\,\Delta
\mu_{\scriptscriptstyle VBF+VH}& \Delta \mu_{\scriptscriptstyle VBF+VH}^2 
      \end{array}\right), \hspace*{0.5cm} \Delta
\mu_i\equiv\sqrt{\left( \Delta\mu_i^{exp} \right)^2 + \left( \Delta\mu_i^{th} \right)^2}\;.
	\label{eq:correlationmatrix}
\end{equation}
The values of $\mu^{exp}_i$, $\Delta\mu^{exp}_i$ and $\rho$ are
extracted from the experimental results, see Appendix
\ref{AppEllipse}. 
The theoretical value $\mu^{th}_1 =
\mu^{th}_{\scriptscriptstyle ggF+tth}$ ($\mu^{th}_2 =
\mu^{th}_{\scriptscriptstyle VBF+VH}$) in the final state channel $X$
is obtained by computing in our model the sum of the ggF and tth (VBF
and VH) production cross sections and multiplying this with the
branching ratio into the final state $X$. Subsequently, the value obtained
is normalized to the corresponding SM rate. The final states that we
take into account are $X=W,Z,\gamma,b$ and $\tau$. The theoretical
uncertainties $\Delta\mu_i^{th}$ stem from 
the scale and PDF uncertainties of the total cross section. We use the
relative theoretical uncertainties of the SM throughout the numerical
analysis, as we checked explicitly for some parameter points that the
theoretical uncertainties obtained within our model are only
slightly modified compared to the SM. This leads then to $\Delta
\mu_{\scriptscriptstyle VBF+VH}^{th}=0$ and very 
small $\Delta \mu_{\scriptscriptstyle ggF+tth}^{th}$. As we computed
all the production cross sections at NLO QCD, the 
uncertainties are the ones given at this order. 
Note also that for the $b\bar{b}$ channel, there is no information
available from ATLAS on the correlation. In this case, we then defined
\begin{equation}
 \chi^2_{h \to b\bar{b}} =\frac{(\mu^{exp}_{b}-\mu^{th}_{b})^2}{(\Delta\mu^{exp}
_b)^2+(\Delta\mu^{th}_b)^2 } \; , 
\label{eq:mub}
\end{equation}
where $\mu_b$ is obtained from the sum of all VBF, VH, ggF and tth
production modes times the branching ratio into $b\bar{b}$ normalized
to the corresponding SM rate.
\s

The constraint from the measured value of the CKM matrix element
$|V_{tb}|$ can be treated in two different ways. Either all points with
$|V_{tb}| > 0.92$ are rejected, or the best fit value quoted by the 
experiments is included in the $\chi^2$ test. The 
CMS collaboration measured the value\footnote{The
measurement does not assume unitarity of the CKM matrix.} to be
\cite{CMSsingletop}  
\begin{equation}
 |V_{tb}^{exp}|=1.02\pm 0.046\;.
\end{equation}
The value of $|V_{tb}^{th}|$ in the model considered in this work is taken
from the $W$ coupling to the top and the bottom quark. For the SM
we assume $|V_{tb}^{th}|=1$. The couplings of all
other SM quarks to the $W$ boson in our model are the same as in the
SM. A $\chi^2$ test 
for the constraint on $V_{tb}$ can therefore be written as 
\begin{equation}
 \chi^2_{V_{tb}} = \frac{(|V_{tb}^{exp}|-|V_{tb}^{th}|)^2}{(\Delta
 V_ { tb}^{exp})^2 }\label{chitot}\;.
\end{equation}

\begin{table}[t]
\centering
\begin{tabular}{|c|c|c|c||c|c|c|}\hline
&\multicolumn{3}{|c||}{$|V_{tb}|>0.92$}&\multicolumn{3}{|c|}{ $|V_{tb}|$ in
$\chi^2$}\\\cline{2-7}
Experiment &$\xi$&$\chi^2/n$&$\chi^2_n$ &$\xi$&
$\chi^2/n$&$\chi^2_n$\\\hline
\multirow{2}{*}{ATLAS}&0.105&8.06/9&0.90&
 0.096&12.34/10&1.23\\\cline{2-7}
& 0.0 &17.54/13&1.35& 0.0 &17.73/14&1.25\\\hline
\multirow{2}{*}{CMS} &0.057&5.22/10&0.52&0.055&6.36/11&0.58\\\cline{2-7}
 &0.0& 9.90/14&0.71& 0.0& 10.09/15&0.67\\\hline
\end{tabular}
\caption{Global $\chi^2$ results for the best fit point taking into
  account EWPT and the Higgs results for ATLAS and CMS,
  respectively: {\it Left:} For parameter points which fulfill
  $|V_{tb}| >0.92$. {\it Right:} When including the measured
  value of $|V_{tb}|$ in the $\chi^2$ test.
  The lines for $\xi=0.0$ list for comparison the SM values. 
  The number of degrees of freedom $n$ are counted naively as the
  difference between the number of observables and the number of
  parameters in the model, and $\chi^2_n \equiv \chi^2/n$.} 
\label{tab:02}
\end{table}

We report in Table~\ref{tab:02} the $\chi^2$ values of the best fit
points for our model and, for comparison, the ones for the SM. They
are given for the two different ways of including the constraint from
$V_{tb}$. The best fit point can be different in both cases. The
global $\chi^2$ is obviously increased when including $V_{tb}$,
although in the SM limit where $|V_{tb}^{th}|=1$ was used, the
change is small. The constraint from $|V_{tb}|$ mainly affects
scenarios with lower masses of the lightest resonance. We distinguish
between the data for the Higgs rates of the two experiments ATLAS
\cite{ATLASdata} and CMS \cite{CMSdata}, as no combination exists so
far. The CMS data turns out to be better described than the ATLAS
data. The best fit points are obtained for values of $\xi\approx 
0.1$ for ATLAS and for $\xi\approx 0.05$ for CMS. In our Composite
Higgs Model their $\chi^2$ is slightly smaller than in the SM, due to the
larger number of free parameters. The value of $\chi^2_n \equiv
\chi^2/n$ gives an estimate of the relative goodness of the fit. Note, 
however, that the counting of the number of degrees of freedom is not
obvious as the SM limit  is reached when $\xi\to 0$ and
$M_{10}\to\infty$, and then the other parameters become
meaningless. \s

\begin{figure}[ht]\hspace{-0.1cm}
   \includegraphics[scale=0.31, angle=-90]{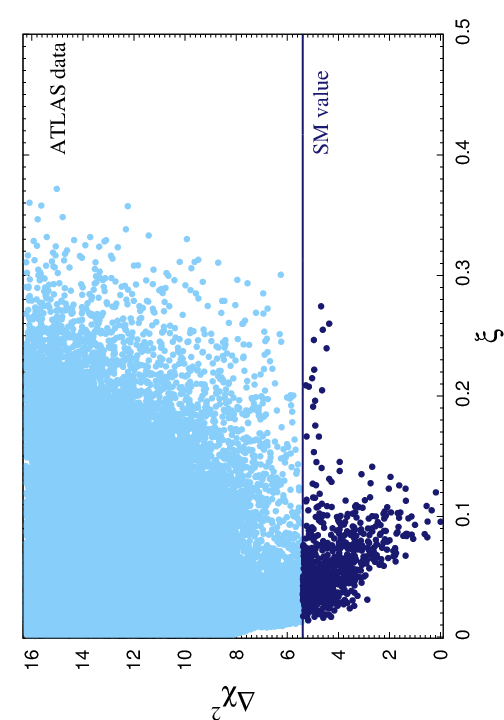}\hspace*{0.5cm}
    \includegraphics[scale=0.30, angle=-90]{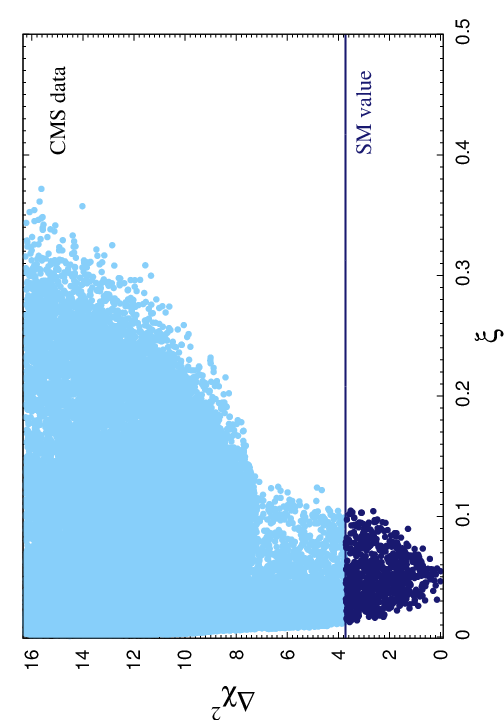}
        \caption{$\Delta \chi^2 = \chi^2-\chi_{min}^2$ taking into
          account the Higgs results of ATLAS (left) and CMS (right),
          as a function of $\xi$. The dark blue points do better than
          the SM, the light blue points have a higher $\Delta\chi^2$.} 
\label{fig:004}
  \end{figure}

\begin{figure}[hb]\hspace{-0.1cm}
   \includegraphics[scale=0.31, angle=-90]{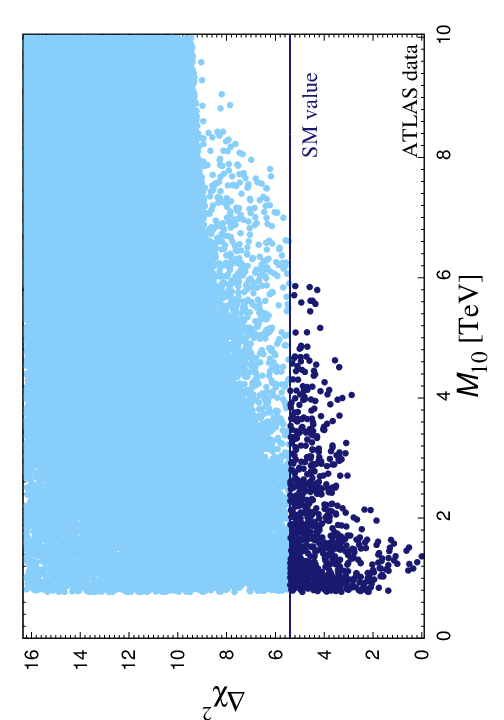}\hspace*{0.5cm}
    \includegraphics[scale=0.30, angle=-90]{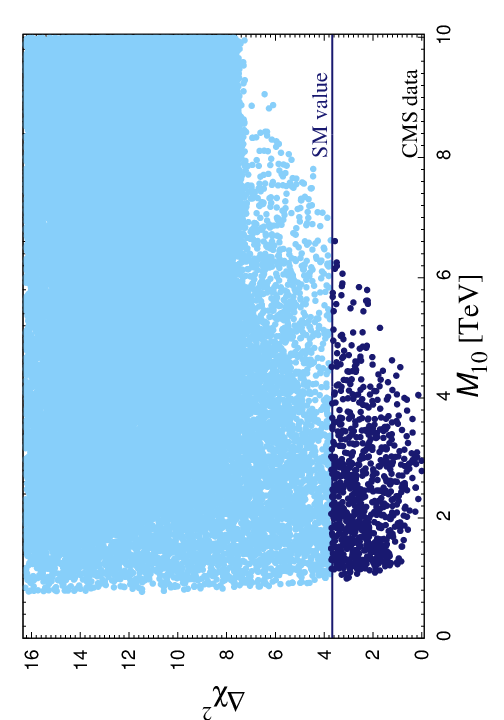}
        \caption{$\Delta \chi^2 = \chi^2-\chi_{min}^2$ taking into
          account the Higgs results of ATLAS (left) and CMS (right),
          as a function of $M_{10}$. The dark blue points do better than
          the SM, the light blue points have a higher $\Delta\chi^2$.} 
\label{fig:expm10}
  \end{figure}

Figure~\ref{fig:004} shows, as a function of $\xi$, $\Delta \chi^2
=\chi^2-\chi_{min}^2$,  where $\chi^2$ is defined in Eq.~(\ref{eq:chialltest}) and
$\chi^2_{min}$ is the value of the best fit point. The color
distinguishes between points which do better than 
the SM and those doing worse. For the CMS results only points with
$\xi\lesssim 0.1$ have a lower $\Delta\chi^2$ than the SM, while for
the ATLAS results this is the case for points up to $\xi\lesssim 0.25$,
although most of the scenarios doing better than the SM are for $\xi
\lsim 0.15$.  Figure~\ref{fig:expm10} shows $\Delta \chi^2$ as a
function of the top and bottom partner mass scale $M_{10}$ for the
ATLAS data (left) and the 
CMS data (right). The lower limit on $M_{10}$ is due to the inclusion of the
direct search bounds on heavy fermion masses. The bulk of the masses
leading to scenarios doing better than the SM lies around
1--2~TeV. This is mainly due to the EWPT. For very heavy fermion
masses the compatibility with the data is not as good. \s

\begin{figure}\hspace{-0.1cm}
   \includegraphics[scale=0.31, angle=-90]{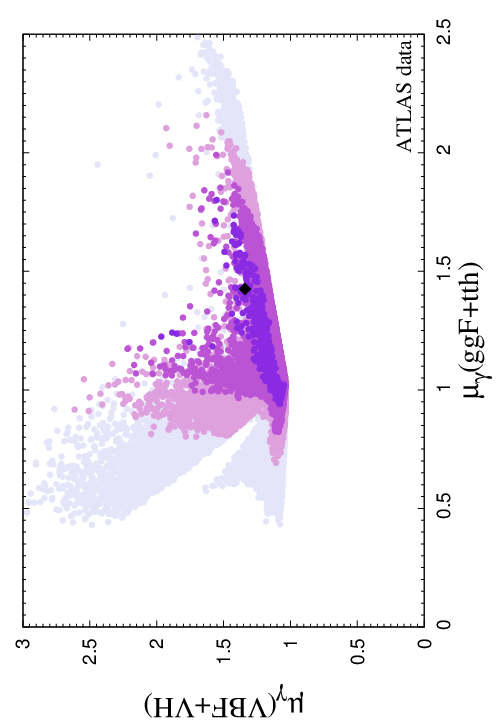}\hspace*{0.5cm}
    \includegraphics[scale=0.30, angle=-90]{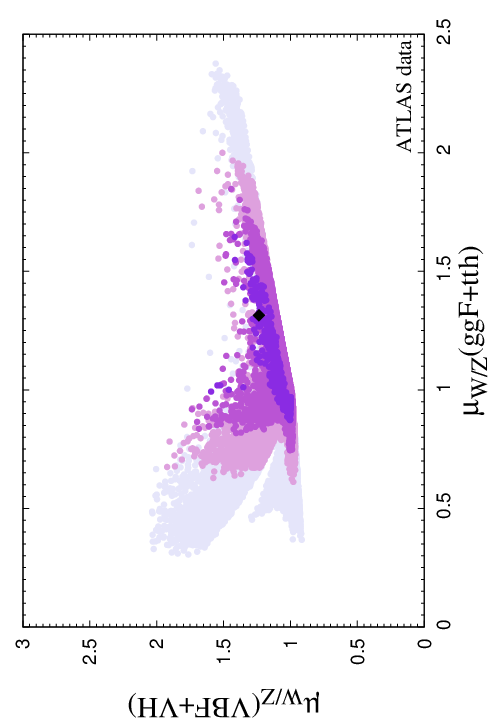}
    \begin{center}
    \includegraphics[scale=0.30, angle=-90]{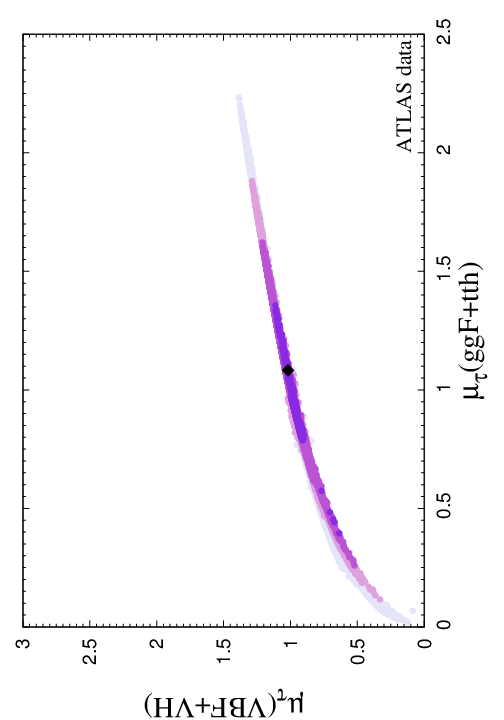}
    \end{center}
        \caption{Fit results obtained from a scan over $\xi$, $y$,
          $\sin \phi_L$ and $M_{10}$ taking into account the EW
          precision data, the measured value of $|V_{tb}|$ and the ATLAS
          Higgs results, shown in the 
$\mu_{\scriptscriptstyle ggF+tth}-\mu_{\scriptscriptstyle VBF+VH}$
plane for the channels $\gamma\gamma$ (top left), 
$W^+W^-$, $ZZ$ (both top right) and $\tau^+\tau^-$ (bottom).
The black rhombus in the plot is the best fit point. The color code in
the plots indicates from dark to light colors the $1\sigma$, $2\sigma$,
$3\sigma$ and $5 \sigma$ regions obtained from the $\chi^2$ test with
four degrees of freedom.}
\label{fig:005}
  \end{figure}
\begin{figure}\hspace{-0.1cm}
   \includegraphics[scale=0.31, angle=-90]{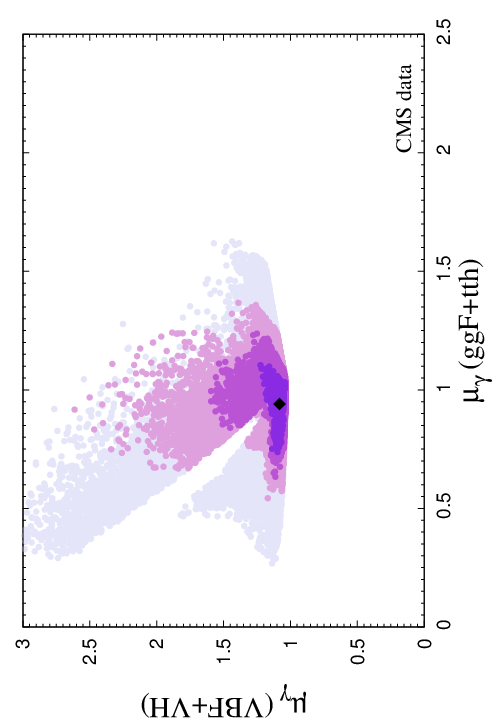}\hspace*{0.5cm}
    \includegraphics[scale=0.30, angle=-90]{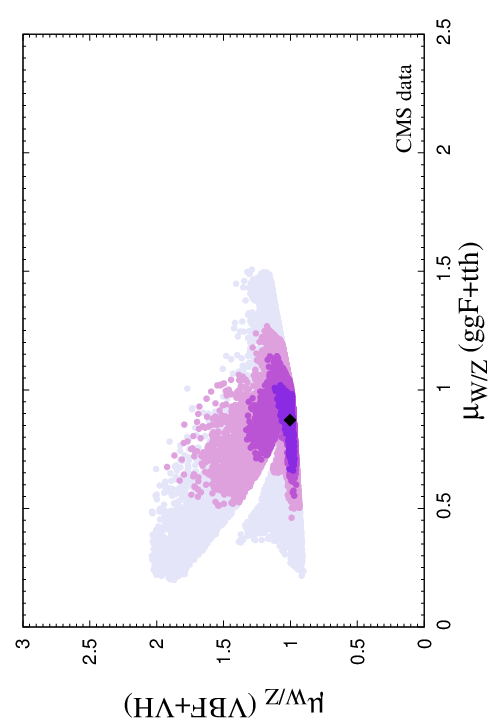}\\
    \includegraphics[scale=0.31, angle=-90]{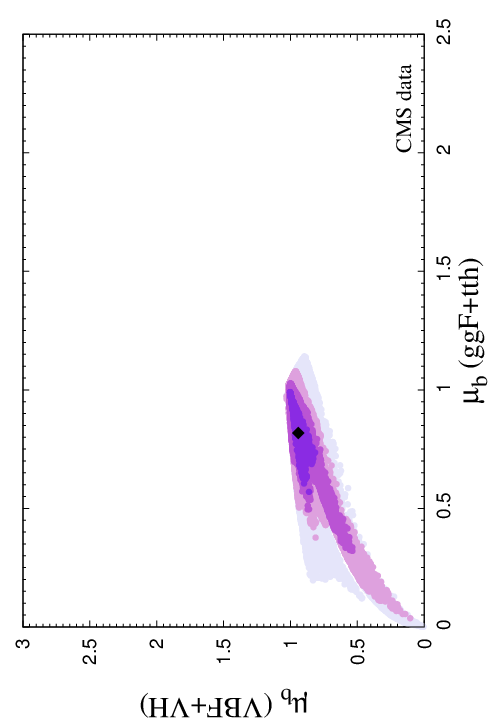}\hspace*{0.5cm}
    \includegraphics[scale=0.30, angle=-90]{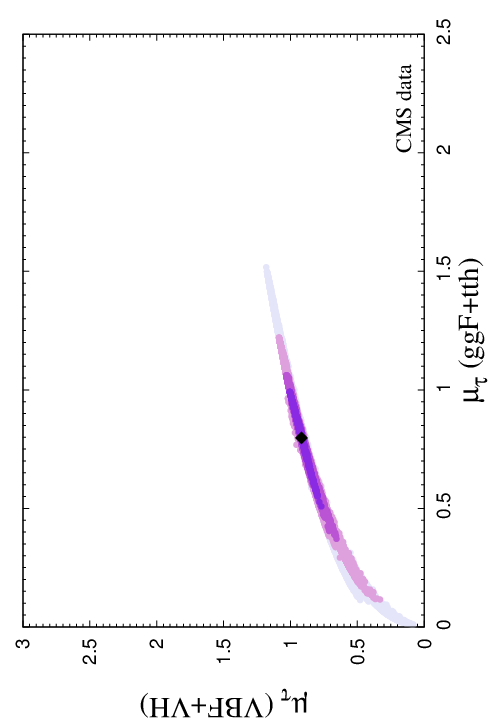}
        \caption{Fit results obtained from a scan over $\xi$, $y$,
          $\sin \phi_L$ and $M_{10}$ taking into account the EW
          precision data, the measured value of $|V_{tb}|$ and the CMS 
          Higgs results, shown in the 
$\mu_{\scriptscriptstyle ggF+tth}-\mu_{\scriptscriptstyle VBF+VH}$ plane for the channels $\gamma\gamma$ (top left),
$W^+W^-$, $ZZ$ (both top right), $b\bar{b}$ (bottom left) and $\tau^+\tau^-$
(bottom right). 
The black rhombus in the plot is the best fit point. The color code in
the plots indicates from dark to light colors the $1\sigma$, $2\sigma$,
$3\sigma$ and $5 \sigma$ regions obtained from the $\chi^2$ test with four degrees of freedom.}
\label{fig:006}
  \end{figure}

In Fig.~\ref{fig:005}, we show the fit results of our parameter scan in the
$\mu_{\scriptscriptstyle ggF+tth}-\mu_{\scriptscriptstyle VBF+VH}$
plane for the Higgs decay channels into $\gamma,W,Z$ and $\tau$
pairs, respectively. The color code indicates from  
dark to light colours the $1\sigma$, $2\sigma$, $3\sigma$ and $5 \sigma$
regions obtained from the $\chi^2$ test as defined in
Eq.~(\ref{eq:chialltest}) 
with the experimental Higgs results reported by ATLAS. The black
rhombus in the plot marks the best fit point which corresponds to the
minimum value obtained from the $\chi^2$ test. The fit contours for
$W$ and $Z$ bosons are the same as their couplings are modified in
the same way due to the custodial symmetry of the model and they are therefore
depicted in the same plot. 
As can be inferred from Fig.~\ref{fig:005} (top left), the ATLAS data
prefer an enhanced Higgs to $\gamma\gamma$ rate.
Also the rate into vector bosons is somewhat enhanced whereas the best
fit point in the $\tau$ channel shows a nearly SM like rate. The same
plots for the CMS Higgs results can be found in
Fig.~\ref{fig:006}, except that additionally the $b\bar{b}$ channel is
shown (bottom left), as CMS provides information about the 
(VBF+VH) and (ggF+tth) production modes and their correlation in the 
$b\bar{b}$ channel. The best fit points are near the SM-like rates in the
$\gamma\gamma$ final state, while the rates in the
$W^+W^-$, $ZZ$, $b\bar{b}$ and $\tau^+\tau^-$ channel are slightly reduced in
the
(ggF+tth) production mode with respect to the SM value. From
Fig.~\ref{fig:005}, bottom, and Fig.~\ref{fig:006}, bottom right, respectively,
we
see that in the $\tau\tau$ final state the region of the points passing the
test is very narrow. In fact this behaviour is already found before
applying the EWPT and $|V_{tb}|$ constraints, {\it i.e.}~the rates for
both production channel combinations behave very similarly. The reason
is that the behaviour of $BR(h\to \tau \tau)$ and of the production in
(ggF+tth) is correlated, and hence the rate $\mu_\tau (VBF+VH)$ is
correlated with the rate $\mu_\tau (ggF+tth)$ via the decay
channel. The former can be easily understood if for the moment the
heavy fermion contributions are left aside (assuming simply the
fermion partners to be very heavy) and the pure Higgs non-linearities
are taken into account. Then both (ggF+tth) production and the decay
into $\tau\tau$ go to zero for $\xi =0.5$ as all the Higgs-Yukawa
couplings are proportional to $(1-2\xi)/\sqrt{1-\xi}$ in this case. With
decreasing $\xi$ from 0.5 to 0 then both the (ggF+tth) production
cross section and the branching ratio ({\it cf.}~Fig.~2 in
\cite{Espinosa:2010vn})  increase. And also the (VBF+VH) production cross
section, which is
proportional to $(1-\xi)$, increases. Due to this strong correlation
between the rates from the two production channel combinations there
remains only a small strip in the $\mu_\tau (ggF+tth)-\mu_\tau
(VBF+VH)$ plane. The effect of imposing the constraints from
EWPT and $|V_{tb}|$ is then to simply divide this strip into 1$\sigma$
to 5$\sigma$ regions. The region in the $b$-quark final 
state, {\it cf.}~Fig.~\ref{fig:006} (bottom left), is explained similarly. It is
somewhat more spread because the Higgs coupling to the bottom quarks
and hence the branching ratio in the $b\bar{b}$ final state is
influenced by the compositeness of the bottom quark. For
the $WW$, $ZZ$ and $\gamma\gamma$ final 
states there is no such strong correlation between the rates, as the
rates from (VBF+VH) production do not vanish for $\xi=0.5$ in this case. \s

So far we have not taken into account the constraint on the mass of
the lightest top partner, as given in
Refs.~\cite{lighttops,MarzoccaSerone,Pomarol:2012qf}. These works
assumed that the Higgs potential is dominated by the first resonances
in the composite sector, and the lightness of the Higgs boson is
related to the lightness of the top partners. An approximate
bound on the mass $m_Q$ of
the lightest top partner was given in Ref.~\cite{Pomarol:2012qf} based
on sum rules: 
\begin{equation}
 m_Q\lesssim \frac{m_h \pi v}{m_t
   \sqrt{N_c}\sqrt{\xi}}\;,\label{lightreslighthiggs} 
\end{equation}
where $N_c = 3$ is the number of colors.\footnote{The formula in
  Eq.~\eqref{lightreslighthiggs} was 
  given for the $\text{MCHM}_5$, but can also be applied for our case, 
  as the mass value, which the lightest resonance can take, is the
  same value for both the {\bf 10} and the {\bf 5} representation,
  see Figure 1 in Ref.~\cite{Pomarol:2012qf}.} This bound eliminates 
automatically large values of $\xi$, as too low masses for the lightest
top partner are already excluded by direct searches. Requiring the
lightest top partner to satisfy Eq.~\eqref{lightreslighthiggs}, the best fit
points are modified compared to Table~\ref{tab:02} and the quality of
the fit becomes slightly worse. The new best fit values for $\xi$ and
$\chi^2$, taking into account this bound, can be found in
Table~\ref{tab:03}. The $\xi$ value for the ATLAS results
becomes somewhat smaller, whereas for the CMS results it hardly
changes. Note, however, that the bound
Eq.~\eqref{lightreslighthiggs} can be relaxed if QCD corrections from a
new heavy gluon of the strong sector are
included~\cite{Barnard:2013hka}. The details depend of course on the
mass of the heavy gluon and its couplings. \s
\begin{table}[t]
\centering
\begin{tabular}{|c|c|c|c|}\hline
Experiment&$\xi$&$m_{t_{lightest}}$&$\chi^2$\\\hline 
ATLAS&
 0.067& 806 GeV&13.71\\\hline
CMS&0.055&1335 GeV&7.17\\\hline
\end{tabular}
\caption{Global $\chi^2$ results for the best fit point respecting
  EWPT, $|V_{tb}|$ and the Higgs search results by ATLAS and 
  CMS, respectively, with the corresponding $\xi$ value and the mass of the
lightest top partner $m_{t_{lightest}}$. In addition
  the constraint of Eq.~\eqref{lightreslighthiggs} originating from
the connection between a light Higgs boson and light resonances has been taken
into account.}
\label{tab:03}
\end{table}

So far we have not discussed the question of fine-tuning in our
model. Experimental data require the electroweak scale $v$ to 
be significantly smaller than the strong symmetry breaking scale $f$. This is
possible through cancellations in the Higgs potential with a precision
that is given by $\Delta = f^2/v^2$. The exact tuning, however,
crucially depends on the actual structure of the Higgs potential,
which in turn is controlled by the choice of the fermion
representations \cite{lighttops,MarzoccaSerone,Pomarol:2012qf}. Therefore
$f^2/v^2 = 1/\xi$ can only be regarded as a measure for the minimal
tuning, while the detailed investigation of the 
amount of fine-tuning of the model would require the calculation of
the Higgs potential. This is beyond the scope of the paper. We therefore
restrict ourselves to state that best compatibility of our 
investigated model with all constraints, that have been taken into
account, is achieved for $\xi$ values around 0.05 which
corresponds to a minimal tuning of $\Delta = 20$. Note that we also
found scenarios with lower $\chi^2$ than in the SM for values of $\xi
\sim 0.3$ which would imply lower tuning. Furthermore, in
composite Higgs models a light Higgs mass can in general only be achieved with
moderate tuning if the mass of the lightest top partner is not too
heavy \cite{lighttops,MarzoccaSerone,Pomarol:2012qf}. With masses for
the lightest top partner of the order of 1~TeV our model can therefore
be estimated to be moderately tuned. 

\section{Conclusions \label{sec:conclusion}}
Composite Higgs Models allow for a smooth deviation from the SM with
identical particle content at low energy. A light narrow Higgs
boson arises as pseudo-Nambu Goldstone boson from the spontaneous
breaking of a strong sector and is separated by a mass gap from the
other resonances of the strong sector. Heavy fermions acquire their
masses by applying the idea of 
partial compositeness: The quark masses are generated through the mixing
with the strong sector by coupling the SM quarks linearly with the heavy
partners of the strong sector. This is in particular interesting for
heavy quarks like the top quark. While in previous investigations the bottom
quark mass has been introduced {\it ad hoc} into the model, we applied in
this work the mass generation through partial compositeness also to
the bottom sector. The model is challenged by strong constraints from the
measurement of the $Zb_L \bar{b}_L$ coupling. The latter is safe from
large corrections only if the $b_L$ belongs to a bi-doublet of
$SU(2)_L \times SU(2)_R$. Starting from a global symmetry group
$SO(5)$, the minimal representation which fulfills this requirement
and incorporates partial compositeness for the bottom quark is
the antisymmetric ${\bf 10}$. Based on a model with the coset
$SO(5)/SO(4)$ and the top and bottom quarks embedded into this
representation, we investigated the phenomenology of Composite Higgs
Models with both top and bottom quarks being partially composite objects. \s

We addressed the constraints due to electroweak
precision measurements. In particular we calculated the loop
corrections to the $Zb_L \bar{b}_L$ coupling due to the heavy top and
bottom partner contributions. The latter did not existed in the
literature before and required the renormalization of the mixing
matrix. Subsequently, we performed a $\chi^2$ test taking into account
EWPT and the recent measurement of $|V_{tb}|$. It turned out that the
fermionic loop contributions drive back the $T$ parameter into the
region compatible with EW precision data, so that the Composite Higgs
Model for some parameter combinations even does better than the SM,
which is not too astonishing in view of the enlarged set of
parameters. The additional contributions from the bottom partners
turned out to have a significant impact on the $\chi^2$ test so that
$\xi$ values of up to 0.2 (0.4) can be obtained at 68\% (99\%) confidence
level, corresponding to a compositeness scale $f$ of 550~GeV (390~GeV). \s

We then proceeded to test the model with respect to its compatibility
with the LHC searches for new heavy fermions and with the 
LHC Higgs search results. For the latter we computed the production cross
sections and branching ratios taking into account the modified Higgs
couplings to the SM particles and the new heavy fermion contributions
in the loop induced processes such as gluon fusion and the decay into
photons.\s

It has been shown before, by applying the low-energy theorem, that
if the determinant of the heavy top mass matrix factorizes into a part
depending on the Higgs non-linearities and a part depending on the
details of the heavy spectrum
-- as it is the case here and in most minimal models --
then the loop-induced Higgs coupling to gluons that
enters the dominant gluon fusion Higgs production process at the
LHC is not sensitive to the details of the spectrum of the top
sector, but only depends on the Higgs non-linearities. In the case of
bottom loops, however, the LET cannot be applied any more, so that the gluon
fusion production cross section now shows a dependence on the masses of
the heavy bottom partners. We performed a global $\chi^2$ test based on the
Higgs signal strengths provided by ATLAS and CMS, on the EWPT and on the
measurement of $|V_{tb}|$. Keeping in addition only those parameter
points which fulfill the limits from the searches for heavy fermions,
we found that numerous scenarios are 
compatible with all the constraints, with the best fit point being
closer to the SM when considering the CMS data than for the ATLAS
data.  For CMS data the best fit point is at $\xi\sim 0.05$, for ATLAS data at
$\xi\sim 0.1$. Seeking for a natural explanation of the
light Higgs boson mass the lightest top partner cannot be too
heavy. Taking this into account the global $\chi^2$ for the best fit point
deteriorates and is now obtained for $\xi \sim 0.07$ for the ATLAS
data, while it hardly changes for the CMS data. The corresponding
lightest top mass here is about 1.3 TeV, for ATLAS data it is around 800 GeV.
\s 

In summary, being guidelined by the principle of introducing a minimum
amount of new parameters, we investigated a Composite Higgs Model with
composite top and bottom quarks. We found that the model is in very
good agreement with the EWPT, the measurement of $V_{tb}$ and the LHC
data from the Higgs and heavy 
fermion searches. Composite bottom partners can even ameliorate the
compatibility of the model with the EWPT. Though the characteristic
scale of the strong sector is pushed to somewhat higher values when 
applying in addition the connection between a light Higgs mass and the
lightest new resonance of the model, it is still in good agreement
with all the constraints.

\subsubsection*{Acknowledgments}

We would like to thank the ATLAS Exotics Group and in particular Mark
Cooke and Merlin Davies for giving us access to the results of ATLAS
searches for top partners. We also want to thank A.~Azatov, A.~Belyaev,
R.~Contino, L.~Di Luzio, M.~Serone, M.~Spira and M.~Wiebusch for
helpful discussions, and C.~Grojean for reading the draft.  R.G. and M.M. are
supported by the DFG/SFB-TR9
{\it Computational Particle 
Physics}. R.G. acknowledges financial support by the {\it
Landesgraduiertenf\"orderung des Landes Baden-W\"urttembergs}.
The CP$^3$-Origins centre is partially funded by the Danish National Research
Foundation, grant number DNRF90.

\section*{Appendix}
\begin{appendix}
\section{The Fermion Couplings to the Gauge Bosons and to the Goldstone bosons}\label{GOLDSTONECOUPLINGS}
For the calculation of the New Physics contributions of our model to
the $Zb_L \bar{b}_L$ coupling we need the couplings of the
fermions to the gauge bosons and to the Goldstone bosons. The former
are obtained from Eq.~\eqref{mod08} after rotation to the mass
eigenstates. The fermion-Goldstone boson couplings have been derived
from the Lagrangian given in Eq.~\eqref{lagragian}, by using
Eq.~\eqref{mod10} and making the identifications according to
Eq.~\eqref{Goldstoneid}. In order to define the couplings in
a general way, the Lagrangians for the specific couplings of the
$W$ bosons, the $Z$ bosons, the charged Goldstone bosons $G^\pm$ and
the neutral Goldstone boson $G^0$ to the quarks $\Psi$ of charge $Q$,
respectively, $Q-1,$ are parameterized as follows 
\begin{align}
 \mathcal{L}_{W}&=\frac{g}{\sqrt{2}}W_{\mu}^+\bar{\Psi}_Q^i
\gamma^{\mu}\left(V_{ij}^{QL}P_L+V_{ij}^{QR}
P_R\right)\Psi_{(Q-1)}^j+h.c.\;,\\
\mathcal{L}_{Z}&=\frac{g}{2 c_W}Z_{\mu}\bar{\Psi}_{Q}^{i}
\gamma^{\mu}\left(X_{ij}^{QL}P_L+X_{ij}^{QR} P_R-2 s_{W}^2
Q\delta_{ij}\right)\Psi_{Q}^j \; ,\\
\mathcal{L}_{G^{\pm}}&=\frac{g}{\sqrt{2}}G^+\bar{\Psi}_Q^i
\left(W_{ij}^{QL}P_L+W_{ij}^{QR}
P_R\right)\Psi_{(Q-1)}^j\;\;\;+h.c\;,\\
\mathcal{L}_{G^0}&=\frac{g}{2 c_W}G^0\bar{\Psi}_Q^i
\left(Y_{ij}^{QL}P_L+Y_{ij}^{QR}
P_R\right)\Psi_{Q}^j\;.&\label{appendix01}
\end{align}
The indices $i,j$ run over the quarks present in the model,
$V^{QL/R},X^{QL/R} ,W^{QL/R}$ and $Y^{QL/R}$ denote the coupling
matrices and $P_{L,R}$ the projectors
\beq
P_{L,R} = \frac{1}{2} (1\mp\gamma_5) \;.
\eeq
Here and in the following we use the abbreviations $c_W \equiv
\cos\theta_W$ and $s_W \equiv \sin\theta_W$. For the coupling of the
$Z$ boson to the quarks we define for later use
\begin{equation}
 \widetilde{X}^{Q, (L,R)}_{ij}\equiv{X}^{Q, (L,R)}_{ij}-2 s_W^2
Q\delta_{ij}\;.\label{appendix02}
\end{equation}
The coupling matrices of the neutral Goldstone boson to the
charge-(-1/3) fermions are given by
\begin{equation}
 Y^{-1/3,L}=i\frac{2 c_W}{g}U_{R}^{b\dagger}\left(\begin{array}{cccc}
 0 & 0 & 0 & 0 \\
 0 & 0 & \frac{y}{2}\sqrt{\xi} & -\frac{y
   \sqrt{1-\xi}}{2 \sqrt{2}} \\
 0 & -\frac{y}{2}\sqrt{\xi} & 0 & -\frac{y
   \sqrt{1-\xi}}{2 \sqrt{2}} \\
 0 & \frac{y \sqrt{1-\xi}}{2 \sqrt{2}} & \frac{y
   \sqrt{1-\xi}}{2 \sqrt{2}} & 0
\end{array}\right)U_{L}^b\;,
\end{equation}
\vspace*{0.1cm}
\begin{equation}
 Y^{-1/3,R}=(Y^{-1/3,L})^{\dagger},
\end{equation}
with $\xi\equiv v^2/f^2$.
And the coupling matrices of the positively charged Goldstone boson to the
charge-2/3 and charge-(-1/3) fermions read
\begin{equation}
W^{2/3,L}=\frac{\sqrt{2}}{g} U_{R}^{t\dagger}
 \left(\begin{array}{cccc}
 0 & 0 & 0 & 0 \\
 0 & 0 & \frac{y}{2}\sqrt{\xi} & -\frac{y
   \sqrt{1-\xi}}{2 \sqrt{2}} \\
 0 & -\frac{y}{2}\sqrt{\xi} & 0 & -\frac{ y
   \sqrt{1-\xi}}{2 \sqrt{2}} \\
 0 & -\frac{1}{2} y \sqrt{1-\xi} & 0 & \frac{y \sqrt{\xi}}{2 \sqrt{2}} \\
 0 & 0 & -\frac{1}{2}  y \sqrt{1-\xi} & -\frac{y \sqrt{\xi}}{2 \sqrt{2}}
\end{array}\right)U_{L}^b\;,
\end{equation}
\begin{equation}
W^{2/3, R}=\frac{\sqrt{2}}{g} U_{L}^{t\dagger} \left(\begin{array}{cccc}
 0 & 0 & 0 & 0 \\
 0 & 0 & \frac{ y}{2}\sqrt{\xi} & -\frac{ y
   \sqrt{1-\xi}}{2 \sqrt{2}} \\
 0 & -\frac{y}{2} \sqrt{\xi} & 0 & -\frac{y
   \sqrt{1-\xi}}{2 \sqrt{2}} \\
 0 & -\frac{1}{2} y \sqrt{1-\xi} & 0 & \frac{y\sqrt{\xi}}{2 \sqrt{2}} \\
 0 & 0 & -\frac{1}{2} y \sqrt{1-\xi} & -\frac{y\sqrt{\xi}}{2 \sqrt{2}}
\end{array}\right) U_{R}^b\;.
\end{equation}

\section{Results for the Corrections to $Zb_L\bar{b}_L$\label{Appendixzbb}}
In this Appendix, the results for the corrections to the decay vertex
$Zb_L\bar{b}_L$ will be presented. The decay
amplitude $\mathcal{M}^{heavy}$ as defined in Eq.~\eqref{epsb02} gets
loop contributions from the top quark and its partners,
$\mathcal{M}^{heavy}_t$, from the bottom quark and its partners,
$\mathcal{M}^{heavy}_b$, and from Higgs bosons in the loops,
$\mathcal{M}^{heavy}_{Higgs}$,
\beq
\mathcal{M}^{heavy} = \mathcal{M}^{heavy}_t + \mathcal{M}^{heavy}_b +
\mathcal{M}^{heavy}_{Higgs} \;.
\eeq
We introduce the reduced masses
\begin{equation}
 y_i=\frac{m_i^2}{m_Z^2}, \hspace*{1cm}
y_W=\frac{m_W^2}{m_Z^2}\hspace*{0.8cm}\text{and}
\hspace*{0.8cm}y_{\beta}^b=\frac{m_{b\beta}^2}{m_Z^2}\;,\label{appendix03}
\end{equation}
where $m_i$ is the mass of one of the top quarks denoted by the index
$i$ and $m_{b\beta}$ the mass of one of
the bottom quarks, denoted by the index $\beta$. With the definitions 
of the gauge and Goldstone boson couplings in
Appendix~\ref{GOLDSTONECOUPLINGS} we then obtain for the contributions
from the top quark and the heavy top partners ($Q=2/3$),
\begin{equation}
\begin{split}
 \mathcal{M}^{heavy}_t=&-\frac{\alpha}{8 \pi s_W^2} \sum_i \left[\sum_j
V_{jb}^{QL} V_{ib}^{QL\star}(2 \tilde{X}_{ij}^{QR}E_1^{ij}+
\tilde{X}_{ij}^{QL}E_2^{ij})
+W_{jb}^{QL} W_{ib}^{QL\star}(
\tilde{X}_{ij}^{QL}E_1^{ij}+\tilde{X}_{ij}^{QR}E_3^{ij})\right]\\&
+\left[\sum_{\beta}\tilde{X}_{b\beta}^{-1/3,L}\left(\frac{1}{2}\left(V_{
i\beta } ^ { QL\star } V_{ib}^{QL}+V_{i\beta}^{QL}
V_{ib}^{QL\star}\right) (2 E_4^{i\beta}-1)
\right.\right.\\&\left.\left.+\frac12\left(W_{i\beta}^{QL\star}
W_{ib}^{QL}+W_{i\beta}^{QL} W_{ib}^{QL\star}\right)
E_4^{i\beta}\right)\right]\\&
+(2 s_W^2-1)\left| W_{ib}^{QL} \right|^2 E_5^i-2 c_W^2 \left| V_{ib}^{QL}
\right|^2 E_6^i+4 s_W^2 
\operatorname{Re}(V_{ib}^{QL \star} W_{ib}^{QL}) E_7^i\\&
-  \sum_{\beta} \tilde{X}_{\beta b }^{-1/3,L} (W_{ib}^{QR\star}
W_{i\beta}^{QL}-4
V_{ib}^{QR} V_{i\beta}^{QL\star})E_8^{i\beta}\;,\label{appendix04}
\end{split}
\end{equation}
where the summation over $i, j$ is over all indices appearing in the top mass
matrix and the summation over $\beta$ over all indices appearing in
the bottom mass matrix. The index $b$ stands for the mass eigenstate
with the bottom quark mass. The abbreviations introduced in the above
formula are given by  
\begin{eqnarray}
E_1^{ij} & = & \sqrt{y_i y_j}\, I_1 (y_i,y_W,y_j)\label{appendix06}\;, 
\\
E_2^{ij} & = & \mbox{Div}-2+y_i+y_j-2 y_W  + 2 I_1 (y_i,y_W,y_j)  \left(y_i-y_W
- 1 \right) \left(y_j-y_W-1
\right) 
\nonumber \\ 
& &-I_2(y_i,y_j) \left(y_i+y_j-2
y_W-3 \right) + \log(y_i) \left(\frac{2y_i}{y_i-y_W}-y_i\right)   
\nonumber \\
& & + \log(y_j) \left(\frac{2 y_j}{y_j-y_W}-y_j\right) + 2 y_W
\log(y_W)\left(1-\frac{ y_i+y_j-2 y_W}{(y_i-y_W)(y_j-y_W)} \right)\;,
\\
E_3^{ij} & = & \frac{1}{2} \Big[ \mbox{Div} + 1 + y_i+y_j-2 y_W + 2 I_1
(y_i,y_W,y_j)  \left(y_i - y_W \right)
\left(y_j - y_W \right)  \nonumber\\ 
& & - I_2(y_i,y_j)
\left(y_i+y_j-2 y_W+1 \right) - y_i \log(y_i) - y_j \log(y_j)  
 + 2 y_W \log(y_W) \Big]\;, 
\end{eqnarray}
\begin{equation}
\begin{split}
 E_4^{i\beta}=\frac{1}{2}\begin{cases}
\text{Re}\left[ -\mbox{Div}+2-\log(y_W)+x_{+}(y^b_{\beta}
  , y_W,y_i)
 \log(1-1/x_{+}(y^b_{\beta},y_W, y_i))\right.\\\left.+ x_{-}(y^b_{\beta},y_W,
 y_i)\log(1-1/x_{-}(y^b_{\beta},y_W,
 y_i))-\frac{y_W-y_i}{y_i}\sqrt{\frac{y^b_{\beta}}{y_i}}E_8^{i\beta}\right]
&\mbox{for }
 y^b_{\beta}\ne 0\;,
 \\ \\
-\mbox{Div}+1-\frac{y_i}{y_i-y_W}
\log(y_i)-\frac{y_W}{y_W-y_i}\log(y_W)+\frac{y_i+y_W}{2(y_i-y_w)
}\\ -\frac { y_i
y_W}{(y_i-y_W)^2}\log(y_i/y_W)
&\mbox{for } y^b_{\beta}= 0\;,
\end{cases}
\end{split}
\end{equation}
\begin{eqnarray}
E_5^{i} & = & \frac{\mbox{Div}}{2} - \frac{1}{2} + y_i-y_W - y_i \log(y_i) + y_W
\log(y_W) \nonumber\\
& &  -  I_1 (y_W,y_i,y_W)
\left((y_i-y_W)^2+y_i\right)- I_2
(y_W,y_W)  \left(y_i-y_W+\frac{1}{2}\right) \;,
\\
E_6^{i} & = & 3\;\mbox{Div}-4+2 \left(y_i-y_W \right)- 2 I_1 (y_W,y_i,y_W)
\left( (y_i-y_W)^2+2 y_W \right) \nonumber\\
& &  - I_2 (y_W,y_W) \left(2
y_i-2 y_W-1 \right)   
\nonumber\\
& & + 2 \log(y_i) \left(\frac{2 y_i}{y_i-y_W}-y_i \right) + 2 \log(y_W)
\left(-\frac{2 y_W}{y_i-y_W}+y_W \right)\;,
\\
E_7^{i} & = & \sqrt{y_W y_i}\, I_1 (y_W,y_i,y_W)\;,
\end{eqnarray}
and
\begin{equation}
\begin{split}
 E_8^{i\beta}=\begin{cases}\sqrt{\frac{y_i}{y_{\beta}^b}}\text{Re}\left[1+\frac{
y_i } { y_W-y_i }
\log\left(\frac { y_W } { y_i } \right)+x_{+}(y^b_{\beta}, y_W,
y_i)\log(1-1/x_{+}(y^b_{\beta},y_W, y_i))\right.\\\left.
+x_{-}(y^b_{\beta},y_W, y_i)\log(1-1/x_{-}(y^b_{\beta},y_W,
y_i))\right]&\hspace{-0.5cm}\mbox{for } y_{\beta}^b\ne 0\;,\label{appendix08}
\\
\\
0 &\hspace{-0.5cm}\mbox{for } y_{\beta}^b= 0\;.
\end{cases}
\end{split}
\end{equation}
with
\begin{equation}
 x_{\pm}(y_1,y_2,y_3)=\frac{1}{2}\left(1+\frac{y_3-y_2}{
y_1}\pm\sqrt{\left(1+\frac{y_3-y_2}{y_1}\right)^2-\frac{4 y_3}{y_1}}\right)\;,
\end{equation}
\begin{eqnarray}
I_1 (y_1,y_2,y_3) & = & -\int_0^1
dx\frac{1}{x+y_2-y_3}\log\left[\frac{x y_1+(1-x) y_2}{x y_1+(1-x)
y_3-x(1-x)}\right]\;,\\
I_2 (y_1,y_2) & = & -\int_0^1 dx\log[x y_1+ (1-x) y_2- x(1-x)]\;.
\end{eqnarray}
The symbol ``Div'' in the formulae stands for the divergent part and 
cancels in the end. The expressions $E_1, E_2, E_3, E_5, E_6$
and $E_7$ are the same as the ones obtained in
Ref.~\cite{Anastasiou:2009rv}, whereas due to the mixing 
matrix renormalization expression $E_4$ changed and an additional
contribution corresponding to the $E_8$ term was added. Note that the
gauge boson self-interactions and the interactions of the Goldstone bosons with
the gauge bosons in the derivation of the result for
$\mathcal{M}^{heavy}_t$ are those of the SM and defined as in
Ref.~\cite{Anastasiou:2009rv}. \s 

In case the fermions in the loop are the bottom quark and its partners,
the amplitude $\mathcal{M}^{heavy}_b$ is obtained from
Eq.~\eqref{appendix04} for $Q=-1/3$ by taking the first three lines and the
last line and
making there the replacements 
\begin{equation}
 y_W\to 1,\hspace*{0.3cm} y_{i,j}\to y^b_{i,j},\hspace*{0.3cm}
V_{ij}^{Q(L,R)}\to
\frac{1}{\sqrt{2} c_W}\widetilde{X}^{Q(L,R)}_{ij}\hspace*{0.3cm}\mbox{ and }
\hspace*{0.3cm}W_{ij}^{Q(L,R)}\to\frac{1}{\sqrt{2}
  c_W}Y_{ij}^{Q(L,R)}\;.
\label{eq:replacement}
\end{equation}
Additionally, for bottom partners in the loop there are also Higgs
contributions. They read 
\begin{equation}
\begin{split}
 \mathcal{M}^{heavy}_{Higgs}=&-\frac{\alpha}{8 \pi s_w^2} \sum_i \left[\sum_j
\tilde{G}^{hbb\star}_{bj} \tilde{G}^{hbb}_{bi}(
\tilde{X}_{ij}^{-1/3,L}E_1^{ij}+\tilde{X}_{ij}^{-1/3,R}E_3^{ij})- 
\tilde{X}_{jb}^{-1/3,L}
(\tilde{G}_{ib}^{hbb\star}
\tilde{G}_{ji}^{hbb\star})E_8^{ij}\right.
\\&\left.+\tilde{X}_{bj}^{-1/3,L}\frac{E_4^{ij}}{2}\left(
\tilde{G}^{hbb}_{ji}
\tilde{G}^{hbb\star}_{bi}+\tilde{G}^{hbb\star}_{ji} \tilde{G}^{hbb}_{bi}
\right)\right]
+\frac{4 s_W^2}{\sqrt{2} c_W} 
\operatorname{Re}(\tilde{G}^{hbb\star}_{bi} X_{ib}^{-1/3,L\star}) E_7^i\; ,
\label{appendix05}
\end{split}
\end{equation}
where in the $E_i$ expressions as given by
Eqs.~\eqref{appendix06}--\eqref{appendix08} the replacements $y_W\to
m_h^2/m_Z^2$ and $y_i\to y^b_i$ have to be done. All summations $i$ and
$j$ are understood as summations over the bottom indices. And we defined
\begin{equation}
 \tilde{G}^{hbb}=\frac{\sqrt{2} s_W}{e}(U^b_L)^{\dagger}G_{hb\bar{b}}U^b_R\;,
\end{equation}
with $U^b_{L,R}$ and $G_{hb\bar{b}}$ as in Eqs.~\eqref{mod09} and
\eqref{mod12}. 
For the SM result $\mathcal{M}^{t+b}_{SM}$, the top-loop contribution
$\mathcal{M}^{t}_{SM}$ has been calculated from Eq.~\eqref{appendix04} by
replacing the couplings with the corresponding SM couplings and by taking
into account only top contributions, {\it i.e.}~no summation over
heavy top partner contributions is performed. Analogously the
bottom-loop contribution $\mathcal{M}^{b}_{SM}$ is obtained from the
first three lines of Eq.~\eqref{appendix04} after making the replacements
Eq.~\eqref{eq:replacement} and by substituting the corresponding SM
couplings where necessary and not taking into account any heavy bottom
partner loops. 
\end{appendix}

\section{Correlation in the Higgs Production Channels}\label{AppEllipse} 

\begin{figure}
	\centering
	\includegraphics[width=0.45\linewidth]{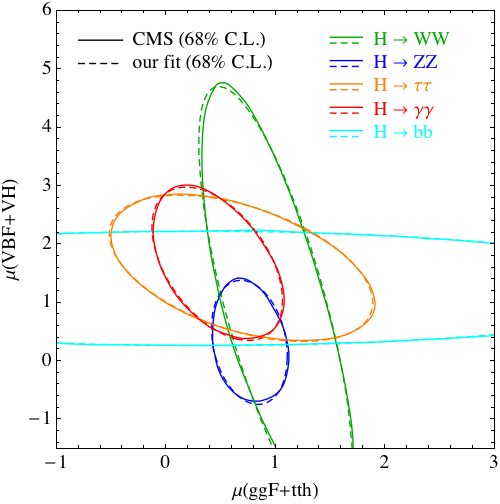}\hspace*{0.5cm}
	\includegraphics[width=0.45\linewidth]{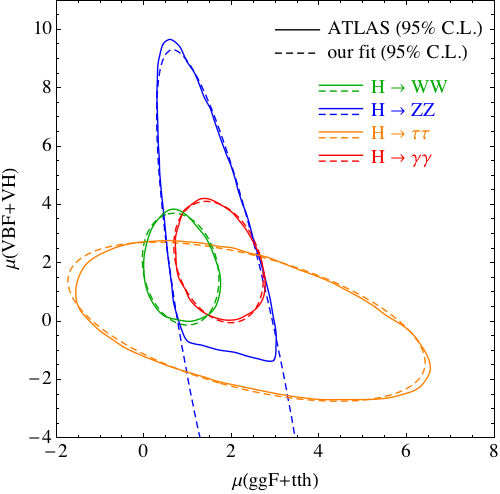}\hspace*{0.5cm}
	\caption{Contours obtained from the experimental
          collaborations~\cite{ATLASdata, CMSdata} (solid lines) and
          from our $\chi^2$ test with two variables (dashed lines) for
          CMS (left, 68\% C.L.) and ATLAS (right, 95\% C.L.) and for
          each Higgs decay channel separately.} 
	\label{fig:ellipses}
\end{figure}

In their measurements of the signal strengths $\mu_i$ for Higgs boson
production and decay, ATLAS and CMS can discriminate between the
different Higgs production mechanisms by looking at the collider
signature of individual events. It is particularly interesting to
separate the production mechanisms involving the coupling of the Higgs
boson to gauge bosons -- vector boson fusion and Higgs-strahlung --
from those involving the coupling of the Higgs boson to fermions --
gluon fusion and associated production with top quarks. The
corresponding signal strengths in a given decay channel are then denoted by
$\mu(\textrm{VBF}+\textrm{VH})$ and $\mu(ggF+tth)$, respectively. The
categorization of a single event into one of the two production
channel combinations, $\mu(\textrm{VBF}+\textrm{VH})$ or
$\mu(ggF+tth)$, is nevertheless ambiguous, and there is therefore an
important correlation among both signal strengths for each decay channel. Both
ATLAS~\cite{ATLASdata} and CMS~\cite{CMSdata} make this correlation
explicit by plotting the 68\% (ATLAS and CMS) and 95\% (ATLAS only)
confidence level contour in the plane
$\mu(\textrm{VBF}+\textrm{VH})-\mu(ggF+tth)$. These contours are
reproduced here in Fig.~\ref{fig:ellipses} (solid lines). The complete 
statistic tests used by the collaborations to produce these contours
are not publicly available, but since the contours  follow obviously
an ellipsoidal shape, we can fit them with the ellipses obtained from
a $\chi^2$ test with two variables. Using the correlation matrix of
Eq.~\eqref{eq:correlationmatrix}, we find for each channel the set of
five parameters $(\mu_{\scriptscriptstyle ggF+tth},
\mu_{\scriptscriptstyle VBF+VH}, \Delta\mu_{\scriptscriptstyle
  ggF+tth}, \Delta\mu_{\scriptscriptstyle VBF+VH}, \rho)$ that give
the best fit between the contour provided by the experiments and the
$\chi^2$ test. The numbers that we obtain are given in
Table~\ref{tab:ellipses}. For CMS, the fit to the 68\% C.L.~contours
matches perfectly. For ATLAS, we choose to fit the 95\% C.L.~contours,
and the agreement is very good as well, although less precise. The
channel $H \to ZZ$  
for ATLAS is peculiar, since the given contour displays a sharp cutoff
for negative values of $\mu(\textrm{VBF}+\textrm{VH})$. Since such
negative values are never reached in our model, the fit given by the
ellipse is fine for our purposes. Notice also that ATLAS does not show
a contour for the channel $H\to b\bar{b}$. Here we use instead the
total signal strength in all production channels, Eq.~\eqref{eq:mub}. 
\begin{table}
	\centering
	\begin{tabular}{|c|c|c|c|c|c|c|}
	\hline
	\multicolumn{2}{|c|}{} & $\mu_{\scriptscriptstyle ggF+tth}$ & $\mu_{\scriptscriptstyle VBF+VH}$ & $\Delta\mu_{\scriptscriptstyle ggF+tth}$ & $\Delta\mu_{\scriptscriptstyle VBF+VH}$ & $\rho$ \\
	\hline
	CMS & $H \to WW$ & 0.761 & 0.321 & 0.229 & 0.701 & -0.226 \\
	& $H \to ZZ$ & 1.001 & 0.944 & 0.464 & 2.481 & -0.739 \\
	& $H \to bb$ & 0.308 & 1.590 & 0.794 & 0.827 & -0.467 \\
	& $H \to \tau\tau$ & 0.684 & 1.591 & 0.794 & 0.827 & -0.467 \\
	& $H \to \gamma\gamma$ & 0.466 & 1.668 & 0.394 & 0.866 & -0.478 \\
	\hline
	ATLAS & $H \to WW$ & 0.828 & 1.796 & 0.358 & 0.782 & -0.178 \\
	& $H \to ZZ$ & 2.119 & -2.132 & 0.751 & 4.679 & -0.800 \\
	& $H \to \tau\tau$ & 2.335 & -0.005 & 1.668 & 1.114 & -0.512 \\
	& $H \to \gamma\gamma$ & 1.695 & 2.041 & 0.418 & 0.849 & -0.273 \\
	\hline
	\end{tabular}
	\caption{Best fit values of the set of parameters
          $(\mu_{\scriptscriptstyle ggF+tth}, \mu_{\scriptscriptstyle
            VBF+VH}, \Delta\mu_{\scriptscriptstyle ggF+tth},
          \Delta\mu_{\scriptscriptstyle VBF+VH}, \rho)$ that reproduce
          the contours provided by ATLAS (at 95\% C.L.) and CMS (at 68\%
          C.L.) for each Higgs boson decay channel, see
Fig.~\ref{fig:ellipses}.}
	\label{tab:ellipses}
\end{table}

\newpage


\begin{thebibliography}{99}

\bibitem{:2012gk}
G.~Aad {\it et al.}  [ATLAS Collaboration],
  Phys.\ Lett.\ B {\bf 716} (2012) 1
  [arXiv:1207.7214 [hep-ex]];
G.~Aad {\it et al.}  [ATLAS Collaboration], ATLAS-CONF-2012-162.

\bibitem{:2012gu}
S.~Chatrchyan {\it et al.}  [CMS Collaboration],
  Phys.\ Lett.\ B {\bf 716} (2012) 30
  [arXiv:1207.7235 [hep-ex]];
S.~Chatrchyan {\it et al.}  [CMS Collaboration], CMS-PAS-HIG-12-045.

\bibitem{strong1} D. B. Kaplan and H. Georgi, Phys. Lett. B {\bf 136}
(1984) 183.

\bibitem{strong2} S. Dimopoulos and J. Preskill, Nucl. Phys. B {\bf
    199} (19829 206; T. Banks, Nucl. Phys. B {\bf 243} (1984) 125;
  D. B. Kaplan, H. Georgi and S. Dimopoulos, Phys. Lett. B {\bf 136}
  (1984) 187; H. Georgi, D. B. Kaplan and P. Galison, Phys. Lett. B
  {\bf 143} (1984) 152; H. Georgi and D. B. Kaplan, Phys. Lett. B {\bf
    145} (1984) 216; M. J. Dugan, H. Georgi and D. B. Kaplan,
  Nucl. Phys. B {\bf 254} (1984) 299.

\bibitem{Giudice:2007fh}
  G.~F.~Giudice, C.~Grojean, A.~Pomarol and R.~Rattazzi,
  JHEP {\bf 0706} (2007) 045
  [hep-ph/0703164].
 
\bibitem{partial}
R. Contino, T. Kramer, M. Son, R. Sundrum, JHEP {\bf 0705} (2007)
074 [hep-ph/0612180]; D. B. Kaplan, Nucl. Phys. B {\bf 365} (1991) 259.

\bibitem{Contino:2006qr}
  R.~Contino, L.~Da Rold and A.~Pomarol,
  Phys.\ Rev.\ D {\bf 75} (2007) 055014
  [hep-ph/0612048].

\bibitem{lighttops}
O.~Matsedonskyi, G.~Panico and A.~Wulzer,
  JHEP {\bf 1301} (2013) 164
  [arXiv:1204.6333 [hep-ph]];
M.~Redi and A.~Tesi,
  JHEP {\bf 1210} (2012) 166
  [arXiv:1205.0232 [hep-ph]];
G.~Panico, M.~Redi, A.~Tesi and A.~Wulzer,
  arXiv:1210.7114 [hep-ph];
  D.~Pappadopulo, A.~Thamm and R.~Torre,
  JHEP {\bf 1307} (2013) 058
  [arXiv:1303.3062 [hep-ph]].
  
\bibitem{MarzoccaSerone}
 D.~Marzocca, M.~Serone and J.~Shu,
  JHEP {\bf 1208} (2012) 013
  [arXiv:1205.0770 [hep-ph]].  

\bibitem{Pomarol:2012qf}
   A.~Pomarol and F.~Riva,
  JHEP {\bf 1208} (2012) 135
  [arXiv:1205.6434 [hep-ph]].

\bibitem{Barnard:2013hka}
J.~Barnard, T.~Gherghetta, A.~Medina and T.~S.~Ray,
  JHEP {\bf 1310} (2013) 055
  [arXiv:1307.4778 [hep-ph]].

\bibitem{vlsearch}
  C.~Dennis, M.~Karagoz, G.~Servant and J.~Tseng,
  hep-ph/0701158;
R.~Contino and G.~Servant,
  JHEP {\bf 0806} (2008) 026
  [arXiv:0801.1679 [hep-ph]];
  J.~A.~Aguilar-Saavedra,
  JHEP {\bf 0911} (2009) 030
  [arXiv:0907.3155 [hep-ph]];
  J.~Mrazek and A.~Wulzer,
  Phys.\ Rev.\ D {\bf 81} (2010) 075006
  [arXiv:0909.3977 [hep-ph]];
  G.~Dissertori, E.~Furlan, F.~Moortgat and P.~Nef,
  JHEP {\bf 1009} (2010) 019
  [arXiv:1005.4414 [hep-ph]];
G.~Cacciapaglia, A.~Deandrea, L.~Panizzi, N.~Gaur, D.~Harada and Y.~Okada,
  JHEP {\bf 1203} (2012) 070
  [arXiv:1108.6329 [hep-ph]];
R.~Barcelo, A.~Carmona, M.~Chala, M.~Masip and J.~Santiago,
  Nucl.\ Phys.\ B {\bf 857} (2012) 172
  [arXiv:1110.5914 [hep-ph]];
 K.~Harigaya, S.~Matsumoto, M.~M.~Nojiri and K.~Tobioka,
  arXiv:1204.2317 [hep-ph];
  A.~Azatov {\it et al.},
  arXiv:1204.0455 [hep-ph];
  N.~Vignaroli,
  arXiv:1204.0468 [hep-ph];
  J.~Berger, J.~Hubisz and M.~Perelstein,
  JHEP {\bf 1207} (2012) 016  [arXiv:1205.0013 [hep-ph]];  
A.~Carmona, M.~Chala and J.~Santiago,
  JHEP {\bf 1207} (2012) 049
  [arXiv:1205.2378 [hep-ph]];
N.~Vignaroli,
  Phys.\ Rev.\ D {\bf 86} (2012) 075017
  [arXiv:1207.0830 [hep-ph]];
Y.~Okada and L.~Panizzi,
  Adv.\ High Energy Phys.\  {\bf 2013} (2013) 364936
  [arXiv:1207.5607 [hep-ph]];
A.~De Simone, O.~Matsedonskyi, R.~Rattazzi and A.~Wulzer,
  JHEP {\bf 1304} (2013) 004
  [arXiv:1211.5663 [hep-ph]];
M.~Chala and J.~Santiago,
  Phys.\ Rev.\ D {\bf 88} (2013) 035010
  [arXiv:1305.1940 [hep-ph]];
M.~Redi, V.~Sanz, M.~de Vries and A.~Weiler,
  JHEP {\bf 1308} (2013) 008
  [arXiv:1305.3818 [hep-ph]];
J.~Li, D.~Liu and J.~Shu,
  arXiv:1306.5841 [hep-ph];
A.~Azatov, M.~Salvarezza, M.~Son and M.~Spannowsky,
  arXiv:1308.6601 [hep-ph].
  
\bibitem{Bini:2011zb}
  C.~Bini, R.~Contino and N.~Vignaroli,
  JHEP {\bf 1201} (2012) 157
  [arXiv:1110.6058 [hep-ph]].

\bibitem{Andeen}
  T.~Andeen, C.~Bernard, K.~Black, T.~Childres, L.~Dell'Asta and
N.~Vignaroli,
  arXiv:1309.1888 [hep-ph].
  
\bibitem{Agashe:2004rs}
  K.~Agashe, R.~Contino and A.~Pomarol,
  Nucl.\ Phys.\ B {\bf 719} (2005) 165
  [hep-ph/0412089].
  
\bibitem{Agashe:2005dk}
  K.~Agashe and R.~Contino,
  Nucl.\ Phys.\ B {\bf 742} (2006) 59
  [hep-ph/0510164].

\bibitem{Barbieri:2007bh}
  R.~Barbieri, B.~Bellazzini, V.~S.~Rychkov and A.~Varagnolo,
  Phys.\ Rev.\ D {\bf 76} (2007) 115008
  [arXiv:0706.0432 [hep-ph]].
  
\bibitem{Pomarol:2008bh}
  A.~Pomarol and J.~Serra,
  Phys.\ Rev.\ D {\bf 78} (2008) 074026
  [arXiv:0806.3247 [hep-ph]].
  
\bibitem{Agashe:2006at}
  K.~Agashe, R.~Contino, L.~Da Rold and A.~Pomarol,
  Phys.\ Lett.\ B {\bf 641} (2006) 62
  [hep-ph/0605341].
  
\bibitem{Lodone:2008yy}
  P.~Lodone,
  JHEP {\bf 0812} (2008) 029
  [arXiv:0806.1472 [hep-ph]].

\bibitem{Gillioz:2008hs}
  M.~Gillioz,
  Phys.\ Rev.\ D {\bf 80} (2009) 055003
  [arXiv:0806.3450 [hep-ph]].
  
\bibitem{Anastasiou:2009rv}
  C.~Anastasiou, E.~Furlan and J.~Santiago,
  Phys.\ Rev.\ D {\bf 79} (2009) 075003
  [arXiv:0901.2117 [hep-ph]].

\bibitem{anarchic}
Y. Grossman and M. Neubert, Phys. Lett. B {\bf 474} (2000) 361
[hep-ph/9912408]; 
T. Gherghetta and A. Pomarol, Nucl. Phys. B {\bf 586} (2000) 141
[hep-ph/0003129]; 
S. J. Huber and Q. Shafi, Phys. Lett. B 498, 256 (2001) [hep-ph/0010195].

\bibitem{Csaki:2008zd}
  C.~Csaki, A.~Falkowski and A.~Weiler,
  JHEP {\bf 0809} (2008) 008
  [arXiv:0804.1954 [hep-ph]].
 
\bibitem{u1h3}
A. L. Fitzpatrick, G. Perez and L. Randall, Phys. Rev. Lett. {\bf 100}
(2008) 171604 [arXiv:0710.1869 [hep-ph]];
C. Csaki, A. Falkowski and A. Weiler, Phys. Rev. D {\bf 80} (2009)
016001 [arXiv:0806.3757 [hep-ph]];
C. Csaki, G. Perez, Z. Surujon and A. Weiler, Phys. Rev. D {\bf 81}
(2010) 075025 [arXiv:0907.0474 [hep-ph]].

\bibitem{u2approx}
R. S. Chivukula and H. Georgi, Phys. Lett. B {\bf 188} (1987) 99; 
L. J. Hall and L. Randall, Phys. Rev. Lett. {\bf 65} (1990) 2939; 
G. D'Ambrosio, G. F. Giudice, G. Isidori and A. Strumia,
Nucl. Phys. B {\bf 645} (2002) 155 [hep-ph/0207036]; 
A. J. Buras, Acta Phys. Polon. B {\bf 34} (2003) 5615
[hep-ph/0310208]; 
V. Cirigliano, B. Grinstein, G. Isidori and M. B. Wise,
Nucl. Phys. B {\bf 728} (2005) 121 [hep-ph/0507001];
C. Delaunay, O. Gedalia, S. J. Lee, G. Perez and E. Ponton,
Phys. Rev. D {\bf 83} (2011) 115003 [arXiv:1007.0243 [hep-ph]].

\bibitem{Redi:2011zi}
  M.~Redi and A.~Weiler,
  JHEP {\bf 1111} (2011) 108
  [arXiv:1106.6357 [hep-ph]].

\bibitem{Redi:2012uj}
  M.~Redi,
  Eur.\ Phys.\ J.\ C {\bf 72} (2012) 2030
  [arXiv:1203.4220 [hep-ph]].

\bibitem{flavourconstraints}
N.~Vignaroli,
  Phys.\ Rev.\ D {\bf 86} (2012) 115011
  [arXiv:1204.0478 [hep-ph]].

\bibitem{DaRold:2012sz}
  L.~Da Rold, C.~Delaunay, C.~Grojean and G.~Perez,
  JHEP {\bf 1302} (2013) 149
  [arXiv:1208.1499 [hep-ph]].

\bibitem{Barbieri:2012tu}
  R.~Barbieri, D.~Buttazzo, F.~Sala, D.~M.~Straub and A.~Tesi,
  JHEP {\bf 1305} (2013) 069
  [arXiv:1211.5085 [hep-ph]].

\bibitem{Contino:2003ve}
  R.~Contino, Y.~Nomura and A.~Pomarol,
  Nucl.\ Phys.\ B {\bf 671} (2003) 148
  [hep-ph/0306259].

\bibitem{Low:2009di}
  I.~Low, R.~Rattazzi and A.~Vichi,
  JHEP {\bf 1004} (2010) 126
  [arXiv:0907.5413 [hep-ph]].

\bibitem{Contino:2010mh}
  R.~Contino, C.~Grojean, M.~Moretti, F.~Piccinini and R.~Rattazzi,
  JHEP {\bf 1005} (2010) 089
  [arXiv:1002.1011 [hep-ph]].

\bibitem{Espinosa:2010vn}
  J.~R.~Espinosa, C.~Grojean and M.~Muhlleitner,
  JHEP {\bf 1005} (2010) 065
  [arXiv:1003.3251 [hep-ph]].

\bibitem{Low:2010mr}
  I.~Low and A.~Vichi,
  Phys.\ Rev.\ D {\bf 84} (2011) 045019
  [arXiv:1010.2753 [hep-ph]].

\bibitem{Grober:2010yv}
  R.~Grober and M.~Muhlleitner,
  JHEP {\bf 1106} (2011) 020
  [arXiv:1012.1562 [hep-ph]];
R.~Grober and M.~Muhlleitner,
  PoS CORFU {\bf 2011} (2011) 021.

\bibitem{Espinosa:2012qj}
  J.~R.~Espinosa, C.~Grojean and M.~Muhlleitner,
  EPJ Web Conf.\  {\bf 28} (2012) 08004
  [arXiv:1202.1286 [hep-ph]].

\bibitem{Azatov:2012qz}
  A.~Azatov and J.~Galloway,
  Int.\ J.\ Mod.\ Phys.\ A {\bf 28} (2013) 1330004
  [arXiv:1212.1380].

\bibitem{Montull:2013mla}
M.~Montull, F.~Riva, E.~Salvioni and R.~Torre,
  Phys.\ Rev.\ D {\bf 88} (2013) 095006
  [arXiv:1308.0559 [hep-ph]].

\bibitem{Azatov:2013ura}
  A.~Azatov, R.~Contino, A.~Di Iura and J.~Galloway,
  Phys.\ Rev.\ D {\bf 88} (2013) 075019
  [arXiv:1308.2676 [hep-ph]].

\bibitem{Contino:2013gna}
  R.~Contino, C.~Grojean, D.~Pappadopulo, R.~Rattazzi and A.~Thamm,
  arXiv:1309.7038 [hep-ph].

\bibitem{Falkowski:2007hz}
  A.~Falkowski,
  Phys.\ Rev.\ D {\bf 77} (2008) 055018
  [arXiv:0711.0828 [hep-ph]].

\bibitem{Azatov:2011qy}
  A.~Azatov and J.~Galloway,
  Phys.\ Rev.\ D {\bf 85} (2012) 055013
  [arXiv:1110.5646 [hep-ph]].

\bibitem{Gillioz:2012se}
  M.~Gillioz, R.~Grober, C.~Grojean, M.~Muhlleitner and E.~Salvioni,
  JHEP {\bf 1210} (2012) 004
  [arXiv:1206.7120 [hep-ph]].

\bibitem{Delaunay:2013iia}
  C.~Delaunay, C.~Grojean and G.~Perez,
  JHEP {\bf 1309} (2013) 090
  [arXiv:1303.5701 [hep-ph]].

\bibitem{Kniehl:1995tn}
  J.~R.~Ellis, M.~K.~Gaillard and D.~V.~Nanopoulos,
  Nucl.\ Phys.\ B {\bf 106} (1976) 292;
M.~A.~Shifman, A.~I.~Vainshtein, M.~B.~Voloshin and V.~I.~Zakharov,
  Sov.\ J.\ Nucl.\ Phys.\  {\bf 30} (1979) 711
   [Yad.\ Fiz.\  {\bf 30} (1979) 1368];
  B.~A.~Kniehl and M.~Spira,
  Z.\ Phys.\ C {\bf 69} (1995) 77
  [hep-ph/9505225].
  
\bibitem{Barducci:2013wjc}
  D.~Barducci, A.~Belyaev, M.~S.~Brown, S.~De Curtis, S.~Moretti and
G.~M.~Pruna,
  JHEP {\bf 1309} (2013) 047
  [arXiv:1302.2371 [hep-ph]].
  
\bibitem{lightflavour2}
  C.~Delaunay, T.~Flacke, J.~Gonzalez-Fraile, S.~J.~Lee, G.~Panico and G.~Perez,
  arXiv:1311.2072 [hep-ph].
  
\bibitem{Carmona:2013cq}
  A.~Carmona and F.~Goertz,
  JHEP {\bf 1304} (2013) 163
  [arXiv:1301.5856 [hep-ph]].
 
\bibitem{Peskin:1991sw}
  M.~E.~Peskin and T.~Takeuchi,
  Phys.\ Rev.\ D {\bf 46} (1992) 381.
  
\bibitem{Altarelli:1990zd}
 G.~Altarelli and R.~Barbieri,
 Phys.\ Lett.\ B {\bf 253} (1991) 161;
  G.~Altarelli, R.~Barbieri and S.~Jadach,
  Nucl.\ Phys.\ B {\bf 369} (1992) 3
   [Erratum-ibid.\ B {\bf 376} (1992) 444];
 G.~Altarelli, R.~Barbieri and F.~Caravaglios,
  Nucl.\ Phys.\ B {\bf 405} (1993) 3.
  
\bibitem{Lavoura:1992np}
  L.~Lavoura and J.~P.~Silva,
  Phys.\ Rev.\ D {\bf 47} (1993) 2046.

\bibitem{Contino:2010rs}
  R.~Contino,
  arXiv:1005.4269 [hep-ph].
 
\bibitem{Grojean:2013qca}
  C.~Grojean, O.~Matsedonskyi and G.~Panico,
  JHEP {\bf 1310} (2013) 160
  [arXiv:1306.4655 [hep-ph]].

\bibitem{Denner:1991kt}
  A.~Denner,
  Fortsch.\ Phys.\  {\bf 41} (1993) 307
  [arXiv:0709.1075 [hep-ph]].
  
\bibitem{Gonzalez:2011he}
  P.~Gonzalez, J.~Rohrwild and M.~Wiebusch,
  Eur.\ Phys.\ J.\ C {\bf 72} (2012) 2007
  [arXiv:1105.3434 [hep-ph]].

\bibitem{Denner:1990yz}
  A.~Denner and T.~Sack,
  Nucl.\ Phys.\ B {\bf 347} (1990) 203.
  
\bibitem{Gambino:1998ec}
  P.~Gambino, P.~A.~Grassi and F.~Madricardo,
  Phys.\ Lett.\ B {\bf 454} (1999) 98
  [hep-ph/9811470]; 
  B.~A.~Kniehl, F.~Madricardo and M.~Steinhauser,
  Phys.\ Rev.\ D {\bf 62} (2000) 073010
  [hep-ph/0005060];   A.~Barroso, L.~Brucher and R.~Santos,
  Phys.\ Rev.\ D {\bf 62} (2000) 096003
  [hep-ph/0004136].
  
\bibitem{Yamada:2001px}
  Y.~Yamada,
  Phys.\ Rev.\ D {\bf 64} (2001) 036008
  [hep-ph/0103046].
  
\bibitem{Mertig:1990an}
  R.~Mertig, M.~Bohm and A.~Denner,
  Comput.\ Phys.\ Commun.\  {\bf 64} (1991) 345.

 
 \bibitem{Kublbeck:1990xc}
J.~Kublbeck, M.~Bohm and A.~Denner,
   Comput.\ Phys.\ Commun.\  {\bf 60} (1990) 165;
T.~Hahn,
    Comput.\ Phys.\ Commun.\  {\bf 140} (2001) 418
    [hep-ph/0012260].
    
 \bibitem{Hahn:1998yk}   
T.~Hahn and M.~Perez-Victoria,
   Comput.\ Phys.\ Commun.\  {\bf 118} (1999) 153
   [hep-ph/9807565];
T.~Hahn,
    Comput.\ Phys.\ Commun.\  {\bf 178} (2008) 217
    [hep-ph/0611273].
  
    \bibitem{ALEPH:2005ab}
  ALEPH, DELPHI, L3, OPAL, SLD, LEP Electroweak Working Group, SLD Electroweak
Group, SLD Heavy Flavor Group Collaborations,
  Phys.\ Rept.\  {\bf 427} (2006) 257
  [hep-ex/0509008].
 
\bibitem{Aaltonen:2012bp}
  T.~Aaltonen {\it et al.}  [CDF Collaboration],
  Phys.\ Rev.\ Lett.\  {\bf 108} (2012) 151803
  [arXiv:1203.0275 [hep-ex]];
   V.~M.~Abazov {\it et al.}  [D0 Collaboration],
  Phys.\ Rev.\ Lett.\  {\bf 108} (2012) 151804
  [arXiv:1203.0293 [hep-ex]];
   T.~E.~W.~Group [CDF and D0 Collaborations],
  arXiv:1204.0042 [hep-ex].

\bibitem{CMSsingletop}
  S.~Chatrchyan {\it et al.}  [CMS Collaboration],
  JHEP {\bf 1212} (2012) 035
  [arXiv:1209.4533 [hep-ex]].
 
\bibitem{georgi}
H.~M.~Georgi, S.~L.~Glashow, M.~E.~Machacek and D.~V.~Nanopoulos,
  Phys.\ Rev.\ Lett.\  {\bf 40}, 692 (1978).

\bibitem{Dawson:2012di}
  S.~Dawson and E.~Furlan,
  Phys.\ Rev.\ D {\bf 86} (2012) 015021
  [arXiv:1205.4733 [hep-ph]].

\bibitem{Dawson:2012mk}
  S.~Dawson, E.~Furlan and I.~Lewis,
  Phys.\ Rev.\ D {\bf 87} (2013) 014007
  [arXiv:1210.6663 [hep-ph]].

\bibitem{ggnlo}
\underline{\it Including the full mass dependence:}
D. Graudenz, M. Spira and P. Zerwas,
Phys.\ Rev.\ Lett.\ {\bf 70} (1993) 1372;
  M.~Spira, A.~Djouadi, D.~Graudenz and P.~M.~Zerwas,
  Phys.\ Lett.\  {\bf B318} (1993) 347; 
  Nucl.\ Phys.\  {\bf B453} (1995) 17
  [hep-ph/9504378];
R. Harlander and P. Kant,
JHEP {\bf 0512} (2005) 015 [hep-ph/0509189];
\underline{\it Heavy mass limit:}
A. Djouadi, M. Spira and P.M. Zerwas, 
Phys.\ Lett.\ {\bf B264} (1991) 440;
  D.~Graudenz, M.~Spira and P.~M.~Zerwas,
  Phys.\ Rev.\ Lett.\  {\bf 70} (1993) 1372;
  S.~Dawson,
  Nucl.\ Phys.\  B {\bf 359}, 283 (1991);
  R.~P.~Kauffman and W.~Schaffer,
  Phys.\ Rev.\  D {\bf 49}, 551 (1994)
  [hep-ph/9305279];
  S.~Dawson and R.~Kauffman,
  Phys.\ Rev.\  D {\bf 49}, 2298 (1994)
  [hep-ph/9310281];
  M.~Kramer, E.~Laenen and M.~Spira,
  Nucl.\ Phys.\  B {\bf 511}, 523 (1998)
  [hep-ph/9611272].

\bibitem{ggnnlo}
  R.~V.~Harlander and W.~B.~Kilgore,
  Phys.\ Rev.\ Lett.\  {\bf 88} (2002) 201801
  [hep-ph/0201206];
 C.~Anastasiou and K.~Melnikov,
  Nucl.\ Phys.\  B {\bf 646}, 220 (2002)
  [hep-ph/0207004];
 V.~Ravindran, J.~Smith and W.~L.~van Neerven,
  Nucl.\ Phys.\  B {\bf 665} (2003) 325 [hep-ph/0302135];
R. V. Harlander and W. B. Kilgore, 
JHEP {\bf 0210} (2002) 017, arXiv:hep-ph/0208096 [hep-ph];
C. Anastasiou and K. Melnikov, 
Phys. Rev. {\bf D67} (2003) 037501, arXiv:hep-ph/0208115 [hep-ph].

\bibitem{topmasseffect}
 R.~V.~Harlander and K.~J.~Ozeren,
  Phys.\ Lett.\  B {\bf 679}, 467 (2009)
  [hep-ph/0907.2997];
 R.~V.~Harlander and K.~J.~Ozeren,
  JHEP {\bf 0911}, 088 (2009)
  [hep-ph/0909.3420];
 A.~Pak, M.~Rogal and M.~Steinhauser,
  Phys.\ Lett.\  B {\bf 679}, 473 (2009)
  [hep-ph/0907.2998];
 A.~Pak, M.~Rogal and M.~Steinhauser,
  JHEP {\bf 1002} (2010) 025
  [arXiv:0911.4662 [hep-ph]].
  
\bibitem{resum}
  S.~Catani, D.~de Florian, M.~Grazzini and P.~Nason,
  JHEP {\bf 0307}, 028 (2003)
  [hep-ph/0306211];
S. Catani et al., 
JHEP {\bf 07} (2003) 028, hep-ph/0306211;
S. Moch and A. Vogt, 
Phys. Lett. {\bf B631} (2005) 48, hep-ph/0508265;
V. Ravindran, 
Nucl. Phys. {\bf B746} (2006) 58, arXiv:hep-ph/0512249 [hep-ph]
  [hep-ph/0911.4662].

\bibitem{ggnnnlo}
 C.~Anastasiou, S.~Buehler, C.~Duhr and F.~Herzog,
  JHEP {\bf 1211} (2012) 062
  [arXiv:1208.3130 [hep-ph]];
 M.~H\"oschele, J.~Hoff, A.~Pak, M.~Steinhauser and T.~Ueda,
  Phys.\ Lett.\ B {\bf 721} (2013) 244
  [arXiv:1211.6559 [hep-ph]];
 C.~Anastasiou, C.~Duhr, F.~Dulat and B.~Mistlberger,
  JHEP {\bf 1307} (2013) 003
  [arXiv:1302.4379 [hep-ph]];
R.~D.~Ball, M.~Bonvini, S.~Forte, S.~Marzani and G.~Ridolfi,
  arXiv:1303.3590 [hep-ph];
S.~Buehler and A.~Lazopoulos,
  JHEP {\bf 1310} (2013) 096
  [arXiv:1306.2223 [hep-ph]].

  \bibitem{Furlan:2011uq}
  E.~Furlan,
  JHEP {\bf 1110} (2011) 115
  [arXiv:1106.4024 [hep-ph]].
  
\bibitem{Dawson:2013uqa}
  S.~Dawson and E.~Furlan,
  arXiv:1310.7593 [hep-ph].

\bibitem{ewcorr}
 A.~Djouadi and P.~Gambino,
  Phys.\ Rev.\ Lett.\  {\bf 73}, 2528 (1994)
  [hep-ph/9406432];
 A.~Ghinculov and J.~J.~van der Bij,
  Nucl.\ Phys.\  B {\bf 482}, 59 (1996)
  [hep-ph/9511414];
 A.~Djouadi, P.~Gambino and B.~A.~Kniehl,
  Nucl.\ Phys.\  B {\bf 523}, 17 (1998)
  [hep-ph/9712330];
 G.~Degrassi and F.~Maltoni,
  Phys.\ Lett.\  B {\bf 600} (2004) 255;
  [hep-ph/0407249].
 U.~Aglietti, R.~Bonciani, G.~Degrassi and A.~Vicini,
  [hep-ph/0610033];
 S.~Actis, G.~Passarino, C.~Sturm and S.~Uccirati,
  Phys.\ Lett.\  B {\bf 670}, 12 (2008)
  [hep-ph/0809.1301];
 C.~Anastasiou, R.~Boughezal and F.~Petriello,
  JHEP {\bf 0904}, 003 (2009)
  [hep-ph/0811.3458].

\bibitem{higlu}
  M.~Spira,
  hep-ph/9510347;
    M.~Spira,
  Nucl.\ Instrum.\ Meth.\ A {\bf 389} (1997) 357
  [hep-ph/9610350].
  
\bibitem{higluweb}
{\it See in the program code:} URL: {\tt http://people.web.psi.ch/spira/higlu/}.

\bibitem{wzfusion} 
 R.~N.~Cahn and S.~Dawson,
  Phys.\ Lett.\  B {\bf 136}, 196 (1984)
  [Erratum-ibid.\  B {\bf 138}, 464 (1984)];
 K.~I.~Hikasa,
  Phys.\ Lett.\  B {\bf 164}, 385 (1985)
  [Erratum-ibid.\  {\bf 195B}, 623 (1987)];
 G.~Altarelli, B.~Mele and F.~Pitolli,
  Nucl.\ Phys.\  B {\bf 287}, 205 (1987).

\bibitem{vbfnloqcd}
  T.~Han, G.~Valencia and S.~Willenbrock,
  Phys.\ Rev.\ Lett.\  {\bf 69} (1992) 3274
  [hep-ph/9206246];
  T.~Figy, C.~Oleari and D.~Zeppenfeld,
  Phys.\ Rev.\ D {\bf 68} (2003) 073005
  [hep-ph/0306109];
   E.~L.~Berger and J.~M.~Campbell,
  Phys.\ Rev.\ D {\bf 70} (2004) 073011
  [hep-ph/0403194].

 \bibitem{Spira:1997dg}
  M.~Spira,
  Fortsch.\ Phys.\  {\bf 46} (1998) 203
  [hep-ph/9705337].

\bibitem{Bolzoni:2010xr}
  P.~Bolzoni, F.~Maltoni, S.~O.~Moch and M.~Zaro,
  Phys.\ Rev.\ Lett.\  {\bf 105} (2010) 011801
  [arXiv:1003.4451 [hep-ph]];
R.~V.~Harlander, J.~Vollinga and M.~M.~Weber,
  Phys.\ Rev.\ D {\bf 77} (2008) 053010
  [arXiv:0801.3355 [hep-ph]].

  \bibitem{Ciccolini:2007ec}
  M.~Ciccolini, A.~Denner and S.~Dittmaier,
Phys. Rev. Lett. {\bf 99} (2007) 161803 [arXiv:0707.0381 [hep-ph]];
  Phys.\ Rev.\ D {\bf 77} (2008) 013002
  [arXiv:0710.4749 [hep-ph]].

\bibitem{V2HV}
  URL: {\tt http://people.web.psi.ch/spira/proglist.html}
  
  
  \bibitem{V2HVnlo}
 T.~Han and S.~Willenbrock,
 Phys.\ Lett.\ B {\bf 273} (1991) 167.



\bibitem{Hamberg:1990np}
  R.~Hamberg, W.~L.~van Neerven and T.~Matsuura,
  Nucl.\ Phys.\ B {\bf 359} (1991) 343
   [Erratum-ibid.\ B {\bf 644} (2002) 403];
   O.~Brein, A.~Djouadi and R.~Harlander,
  Phys.\ Lett.\ B {\bf 579} (2004) 149
  [hep-ph/0307206].
 
\bibitem{Ciccolini:2003jy}
  M.~L.~Ciccolini, S.~Dittmaier and M.~Kramer,
  Phys.\ Rev.\ D {\bf 68} (2003) 073003
  [hep-ph/0306234].

\bibitem{lotth} 
 R.~Raitio and W.~W.~Wada,
  Phys.\ Rev.\  D {\bf 19}, 941 (1979);
 J.~N.~Ng and P.~Zakarauskas,
  Phys.\ Rev.\  D {\bf 29}, 876 (1984);
 Z.~Kunszt,
  Nucl.\ Phys.\  B {\bf 247}, 339 (1984);
J. F. Gunion, 
Phys.\ Lett.\ B {\bf 261} (1991) 510;
 W.~J.~Marciano and F.~E.~Paige,
  Phys.\ Rev.\ Lett.\  {\bf 66}, 2433 (1991).

\bibitem{ttHnlo}
    W.~Beenakker, S.~Dittmaier, M.~Kramer, B.~Plumper, M.~Spira and
P.~M.~Zerwas,
  Phys.\ Rev.\ Lett.\  {\bf 87} (2001) 201805
  [hep-ph/0107081];
 Nucl.\ Phys.\ B {\bf 653} (2003) 151
  [hep-ph/0211352];
   L.~Reina and S.~Dawson,
  Phys.\ Rev.\ Lett.\  {\bf 87} (2001) 201804
  [hep-ph/0107101];
   S.~Dawson, L.~H.~Orr, L.~Reina and D.~Wackeroth,
  Phys.\ Rev.\ D {\bf 67} (2003) 071503
  [hep-ph/0211438].

\bibitem{LHCcxn}
S.~Dittmaier {\it et al.}  [LHC Higgs Cross Section Working Group Collaboration],
  arXiv:1101.0593 [hep-ph];
  URL: {\tt https://twiki.cern.ch/twiki/bin/view/LHCPhysics/CrossSections}
  
\bibitem{hdecay}
  A.~Djouadi, J.~Kalinowski and M.~Spira,
  Comput.\ Phys.\ Commun.\  {\bf 108} (1998) 56
  [hep-ph/9704448];
   A.~Djouadi, M.~M.~Muhlleitner and M.~Spira,
  Acta Phys.\ Polon.\ B {\bf 38} (2007) 635
  [hep-ph/0609292].
 
\bibitem{Contino:2013kra}
  R.~Contino, M.~Ghezzi, C.~Grojean, M.~Muhlleitner and M.~Spira,
  JHEP {\bf 1307} (2013) 035
  [arXiv:1303.3876 [hep-ph]].

\bibitem{ehdecay}
URL: {\tt http://www.itp.kit.edu/$\sim$maggie/eHDECAY/}.

\bibitem{Alloul:2013naa}
  A.~Alloul, B.~Fuks and V.~Sanz,
  arXiv:1310.5150 [hep-ph].

  \bibitem{directsearchesATLAS1}
   ATLAS Collaboration, 
   ATLAS-CONF-2013-018.
\bibitem{directsearchesATLAS2}
   ATLAS Collaboration,
ATLAS-CONF-2013-051,
ATLAS-CONF-2013-056,
and 
ATLAS-CONF-2013-060.
\bibitem{directsearchesCMS}
  CMS Collaboration,
CMS PAS B2G-12-019. 

\bibitem{Chatrchyan:2013uxa}
  S.~Chatrchyan {\it et al.}  [CMS Collaboration],
  arXiv:1311.7667 [hep-ex].

\bibitem{Bhattacharya:2013poa}
  S.~Bhattacharya [on behalf of the CMS Collaboration],
  arXiv:1310.2299 [hep-ex].

\bibitem{Aguilar-Saavedra:2013qpa}
  J.~A.~Aguilar-Saavedra, R.~Benbrik, S.~Heinemeyer and M.~Perez-Victoria,
  arXiv:1306.0572 [hep-ph].
 
\bibitem{dijetsearch}
G. Aad et al. [ATLAS Collaboration], JHEP {\bf 1301} (2013) 029
[arXiv:1210.1718 [hep-ex]]; 
S.~Chatrchyan {\it et al.}  [CMS Collaboration],
  Phys.\ Rev.\ D {\bf 87} (2013) 052017
  [arXiv:1301.5023 [hep-ex]].

\bibitem{Belyaev:2013ida}
   G.~Cacciapaglia, A.~Deandrea, G.~D.~La Rochelle and J.~-B.~Flament,
  JHEP {\bf 1303} (2013) 029
  [arXiv:1210.8120 [hep-ph]];
G.~Belanger, B.~Dumont, U.~Ellwanger, J.~F.~Gunion and S.~Kraml,
  Phys.\ Rev.\ D {\bf 88} (2013) 075008
  [arXiv:1306.2941 [hep-ph]].
A.~Belyaev, M.~S.~Brown, R.~Foadi and M.~T.~Frandsen,
  arXiv:1309.2097 [hep-ph].

 
\bibitem{ATLASdata}
  ATLAS Collaboration,
  ATLAS-CONF-2013-034.
  
\bibitem{CMSdata}
  CMS Collaboration,
  CMS-PAS-HIG-13-005.
  
\end{thebibliography}
\end{document}